\documentclass[preprint]{article}

\usepackage{microtype}
\usepackage{graphicx}
\usepackage{subfigure}
\usepackage{booktabs} % for professional tables
\usepackage{enumerate}
\usepackage{url}
\usepackage{xcolor}
\usepackage{setspace}
\usepackage{multirow}

\usepackage{hyperref}

\usepackage{icml2026}

\usepackage{amsmath}
\usepackage{amssymb}
\usepackage{mathtools}
\usepackage{amsthm}

\usepackage[capitalize,noabbrev]{cleveref}

\theoremstyle{plain}
\newtheorem{theorem}{Theorem}[section]

\newtheorem{lemma}[theorem]{Lemma}

\theoremstyle{definition}

\newtheorem{assumption}[theorem]{Assumption}
\newtheorem*{remark}{Remark}

\usepackage[textsize=tiny]{todonotes}

\newcommand{\R}{\mathbb{R}}
\newcommand{\Rd}{\mathbb{R}^d}
\newcommand{\E}{\mathbb{E}}
\newcommand{\V}{\mathbb{V}}

\icmltitlerunning{Bayes Posterior Sampling for Large GLMMs}

\begin{document}

\twocolumn[
\icmltitle{Safe, Scalable, and Accurate Bayes Posterior Sampling\\
for Large-Data Generalized Linear Mixed Models}

% It is OKAY to include author information, even for blind
% submissions: the style file will automatically remove it for you
% unless you've provided the [accepted] option to the icml2026
% package.

% List of affiliations: The first argument should be a (short)
% identifier you will use later to specify author affiliations
% Academic affiliations should list Department, University, City, Region, Country
% Industry affiliations should list Company, City, Region, Country

% You can specify symbols, otherwise they are numbered in order.
% Ideally, you should not use this facility. Affiliations will be numbered
% in order of appearance and this is the preferred way.
\icmlsetsymbol{equal}{*}

\begin{icmlauthorlist}
\icmlauthor{Youngsoo Baek}{duke}
\icmlauthor{Samuel I. Berchuck}{duke}
\end{icmlauthorlist}

\icmlaffiliation{duke}{Department of Biostatistics \& Bioinformatics, Duke University, Durham, NC, USA}

\icmlcorrespondingauthor{Samuel I. Berchuck}{sib2@duke.edu}

% You may provide any keywords that you
% find helpful for describing your paper; these are used to populate
% the "keywords" metadata in the PDF but will not be shown in the document
\icmlkeywords{Machine Learning, ICML}

\vskip 0.3in
]

% this must go after the closing bracket ] following \twocolumn[ ...

% This command actually creates the footnote in the first column
% listing the affiliations and the copyright notice.
% The command takes one argument, which is text to display at the start of the footnote.
% The \icmlEqualContribution command is standard text for equal contribution.
% Remove it (just {}) if you do not need this facility.

\printAffiliationsAndNotice{}  % leave blank if no need to mention equal contribution
%\printAffiliationsAndNotice{\icmlEqualContribution} % otherwise use the standard text.

\begin{abstract}
% 4--6 sentences long here
We consider the problem of scalable sampling algorithms to fit Bayesian generalized linear mixed models on large datasets.
Stochastic gradient Langevin dynamics, coupled with smooth re-parameterizations of variance parameters, produces divergent Markov chains and cannot be reliably used for sampling covariance parameters of random effects.
We advocate the use of a mirror Langevin dynamics algorithm, propose the novel {\it stochastic mirror Langevin dynamics} based on data subsampling, and provide concrete guidelines for its use in a Bayesian inference framework.
Based on an explicit Wasserstein distance error bound between the posterior and its algorithmic approximation, we propose a post-processing step that yields an asymptotic, order-wise correct estimation of the posterior variance, eliminating the irreducible posterior variance estimation bias due to subsampling.
Empirical performance of the method is evaluated through simulated experiments and a longitudinal study of pain trajectories in a study of breast cancer survivors.
\end{abstract}

\section{Introduction}

A fundamental computational problem in Bayesian inference is to draw samples that approximate the posterior distribution given data and the model. 
We focus on performing Bayesian inference for a class of models referred to as \emph{generalized linear mixed models} (GLMMs). 
In brief, GLMMs extend generalized linear models (GLM) to situations where the data are structured in groups and regression coefficients can vary by group \cite{gelman2007data}.
Bayesian estimation is attractive because it provides both a direct quantification of uncertainty and a principled way to fit models with complicated, multilayered probability specifications \cite{gelman2013bayesian}.
However, scaling Bayesian inference to large datasets, especially for GLMMs, remains challenging.
Unlike linear regression, hierarchical or not, the marginal likelihood of GLMMs for binary or count response is generally intractable and requires numerical integration over multivariate latent variables.
Modern probabilistic programming software avoids the problem by sampling both the parameters of interest and latent variables using conventional Markov chain Monte Carlo (MCMC) tools;
however, this strategy is infeasible as the number of groups, $n$, grows to be large. 
While there exist feasible variational Bayes-based approximation methods \cite{ormerod2012gaussian},
their approximation quality remains generally unclear for complex multilevel probabilistic models.

A suite of recently proposed methods, called stochastic gradient MCMC (SGMCMC) \cite{nemeth2021stochastic}, appears to resolve the scalability problem. 
Most SGMCMC algorithms are based on discretizing the underlying continuous-time stochastic differential equation (SDE) whose stationary distribution coincides with the targeted posterior. 
However, there are currently no concrete guidelines on how to incorporate SGMCMC into the GLMM setting, which often includes parameters \emph{constrained} to a subset of the Euclidean space, including posterior variances of latent variables.
In fact, we demonstrate that using previously proposed SGMCMC methods by smoothly re-parameterizing constrained parameters can produce wildly divergent Markov chains in finite sampling time and is thus unsafe. 
Furthermore, previous results have demonstrated how the invariant distribution of SGLD has a bias asymptotically non-vanishing in $n$ for a fixed step size \cite{brosse2018promises}. 
The problem of refining SGMCMC to deflate the stochastic gradient variance has been previously studied by \cite{baker2019control}. 
They have proposed the use of control variates, assuming that one can initialize the MCMC chains at a value $O(1/\sqrt{n})$ distance from the global minimizer of $f$.
However, for more complex models such as GLMMs, designing effective control variates remains challenging and far from straightforward. For one, the authors have not considered the issue of algorithmic instability described in Section \ref{sec:motivation}, which cannot be remedied using control variates.
Even disregarding the problem, empirical results suggest that \cite{baker2019control} is not as straightforward to adapt to GLMMs,
where it is not easy to obtain exactly unbiased stochastic gradients,
and that the resulting samples for random effects variance tend to underestimate the true posterior variance.
The authors have previously conducted simulated studies addressing the inaccuracy of posterior variance estimate using control variates in GLMMs.
These materials can be found in Section 4.2 \cite{berchuck2026scalablebayesianinferencegeneralized}.
% of the Supplementary Materials.

We address all mentioned challenges of fitting Bayesian GLMMs by developing a novel stochastic mirror Langevin dynamics (SMLD) algorithm for constrained sampling.
Our contribution is threefold.
First, we transfer the extant literature on mirror Langevin dynamics (MLD) \cite{hsieh2018mirrored,ahn2021efficient,li2022mirror} to inference in large-data hierarchical GLMs, including GLMMs. 
Most of the literature on constrained sampling, including MLD algorithm, has a ``theoretical'' flavor, discussing applications such as sampling Brownian motion paths on a polytope. 
Our work advocates the usage of MLD-based algorithms in sampling the covariance parameters of random effects, providing practical guidelines for a scalable GLMM inference. 
Second, the use of SMLD reaps the benefit of algorithmic safety: unlike existing SGMCMC algorithms, which, when applied to a smooth re-parameterization of constrained parameters, can produce divergent Markov chains, SMLD produces ergodic chains for common GLMM likelihoods.
Third, we develop a rigorous error analysis of the algorithm and propose a post-processing step based on asymptotic justifications. 
The method, unlike designing control variates, is entirely non-intrusive to the iterative sampling algorithm.
This post-processing yields an order-wise correct estimate of the curvature around the mode of the posterior in the number of groups.
For large datasets with many groups, to which SMLD algorithms are applicable, this in turn will yield an accurate estimation of the posterior variance.

The rest of this work is structured as follows.
Section \ref{sec:motivation} studies a simple toy example that motivates the need for a novel sampling method in GLMMs.
Section \ref{sec:model_desc} provides a detailed description of GLMMs and the associated Langevin dynamics-based sampling methods, including the hereby proposed SMLD.
Section \ref{sec:mirror} describes the implementation details, an error analysis, and an additional post-processing step that yields an order-wise correct estimate of the curvature around the maximum likelihood estimator (MLE).
Sections \ref{sec:sim_study} and \ref{sec:real_study} study empirical performance of the algorithm on simulated and real-world datasets.
Section \ref{sec:discussion} concludes the paper.
Appendix includes proofs of our main theorems, supplementary theoretical results, and further tables and visualizations for our data experiments.
% Supplementary Materials include the authors' previous paper proposing a similar idea for unconstrained stochastic MCMC, and codes for simulation studies.

\noindent {\bf Notation.}
We write $[n]=\{1,2,\ldots,n\}$ for every natural number $n$.
For two sequences $a_n,b_n$, $a_n = O(b_n)$ implies that there exists some $C>0$ such that $a_n\leq Cb_n$ for all sufficiently large $n$; $a_n\ll b_n$ means the sequence $a_n/b_n\to 0$, and $a_n\asymp b_n$ means $a_n/b_n\to c\in(0,\infty)$.
Unless otherwise noted, all norms $\|\cdot\|$ are Euclidean norms for $\R^d$-vectors.
For each $v\in\R^d$, we write $v^{\otimes2} := vv^\top$, for a succinct notation.
For a $\R^{d\times d}$-matrix $A$, $\|A\|_{HS}$ is the Hilbert-Schmidt norm ${\rm Tr}(A^\top A)$.
Given two continuous probability distributions $p,q$ on $\R^d$, we abuse the notation and also denote their Lebesgue densities by $p,q$. If $X\sim p$ is a $\R^d$-valued random variable with distribution $p$, then for any $F:\R^d\to\R^{d'}$, 
$\E_p[F(X)]\in\R^{d'}$ and $\V_p[F(X)]\in\R^{d'\times d'}$ are the mean vector and covariance matrix of $F(X)$, respectively;
equivalently, e.g., $\E_{F_\#p}[\tilde{X}]$ for $F_\#p$ the pushforward measure of $p$ under $F$.
The space $\mathcal{P}_k(\R^d)$ is the space of all probability distributions on $\R^d$ with finite $k$-th moments.
Given two probability distributions $p,q\in\mathcal{P}_k$, their $k$-Wasserstein distance $W_k(p,q)$ is defined as $\inf_{\Pi}\sqrt{\E_{\Pi}[\|X-Y\|^k]}$ for every natural number $k$, 
where the infimum is taken over all couplings between $p$ and $q$, i.e., probability distributions of a $\R^d\times\R^d$-valued random variable $(X,Y)$ such that the marginals of $X$ and $Y$ are $p$ and $q$, respectively.

\section{Motivation: Sampling the Log-variance}
\label{sec:motivation}

To motivate the need of a new algorithm for scalable GLMMs, we will consider a toy problem that demonstrates the perils of re-parameterizing positively constrained parameters, which routinely appear in GLMMs as modeling group-level variations.
Suppose $y_1,\ldots,y_n\in\R$ are observations known to be drawn from a mean-zero normal distribution with unknown standard deviation $\sigma > 0$.
We place a standard normal prior on $\sigma$ truncated at the origin, though this choice of prior is not essentially relevant. 

Since there are many algorithms to sample generic $\R^d$-valued posterior distributions, we only need to smoothly re-parameterize the positively constrained $\sigma$. A natural choice is to set $\theta = \log\sigma$.
The posterior density of $\theta$ is proportional to 
$e^{-f(\theta)} =
    \exp(-n\theta -\frac{e^{-2\theta}}{2}\sum_{i=1}^{n}y_i^2 + \theta - \frac{e^{2\theta}}{2})$.
We write by each contribution to $f$ from the $i$-th negative data likelihood as $f_i$, with $f_0$ the negative log-prior density.
The derivative $f'(\theta)$ is then given by
\begin{equation}
    \label{eq:grad_var}
    \sum_{i=0}^{n} f_i'(\theta)  =  \left(n - 1\right) -e^{-2\theta}\sum_{i=1}^{n}y_i^2 + e^{2\theta}.
\end{equation}
We focus our attention to sampling methods based on an associated SDE for a Brownian motion $B_t$:
\begin{equation}
    \label{eq:sde}
    d\theta_t = -f'(\theta_t)dt + \sqrt{2}dB_t.
\end{equation}
This SDE is known to have a strong solution whose stationary distribution has a density $\propto e^{-f(\theta)}$ under fairly general regularity conditions (Theorem 2.1 \cite{roberts1996exponential}). 
A natural discrete approximation is given by the unadjusted Langevin algorithm (ULA) \cite{roberts1996exponential}, whose iterates follow the equation for a constant step size $\epsilon = \epsilon(n) > 0$:
\begin{equation}
    \label{eq:ula}
    \theta_{k+1} = \theta_k - \epsilon f'(\theta_k) + \sqrt{2\epsilon}Z_k,\; k = 0,1,2\cdots.
\end{equation}
For datasets of moderate size, one can correct each ULA iterate to have the targeted posterior as an invariant distribution by adding an accept/reject scheme based on Metropolis-Hastings algorithm. 
For large $n$, both ULA and the Metropolis-Hastings step add computational burden. 
One of the first SGMCMC algorithms proposed to address this difficulty was the stochastic gradient Langevin dynamics (SGLD) \cite{welling2011bayesian}.
In our context, the iterates follow the following equation:
\begin{equation}
    \label{eq:sgld}
    \theta_{k+1} = \theta_k - \epsilon\frac{n}{S} 
    \sum_{i\in\mathcal{S}_k}f'_i(\theta_k) + \sqrt{2\epsilon}Z_k,\; k = 0,1,2\cdots;
\end{equation}
where $\mathcal{S}_k$ is a minibatch, of a fixed cardinality $S>0$, and of indices randomly drawn from $[n]$ with replacement. 
While we have introduced the SDE and the corresponding discretizations in the context of a toy model,
their definitions are generic and can be naturally extended to the case when $\theta\in\R^d,\; d>1$.

\begin{figure}[t]
    \centering
    \includegraphics[width=\linewidth]{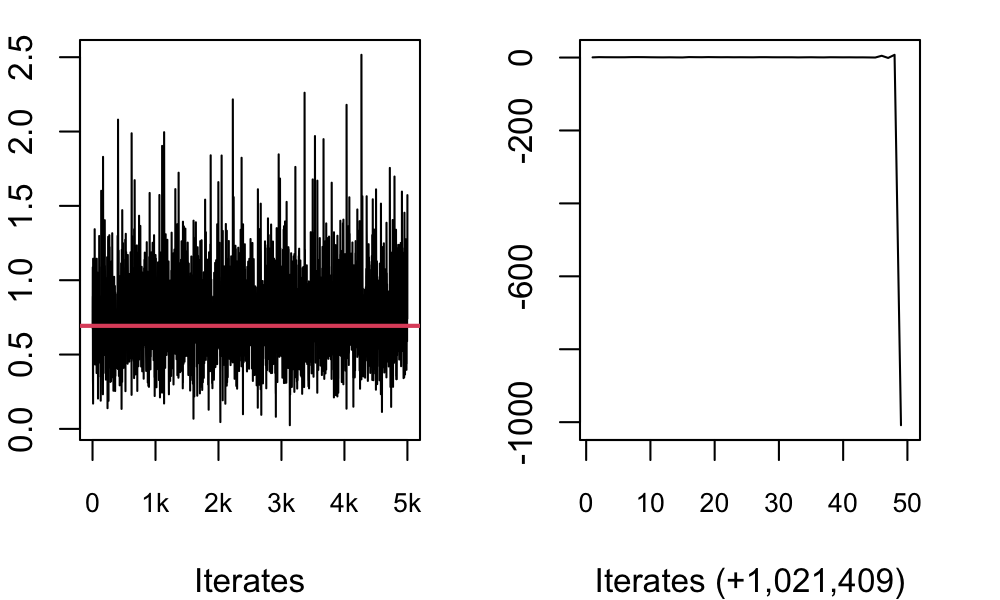}
    \caption{The first 5,000 (left) and the last 50 (right) iterates before numerical divergence of the SGLD iterates run with $f'(\theta)$ given by \eqref{eq:grad_var}. {\color{red} Red line} indicates $\theta_0=\log(2)$ that was used to simulate the dataset.}
    \label{fig:fig1}
\end{figure}

\begin{figure}[t!]
    \centering
    \includegraphics[width=0.75\linewidth]{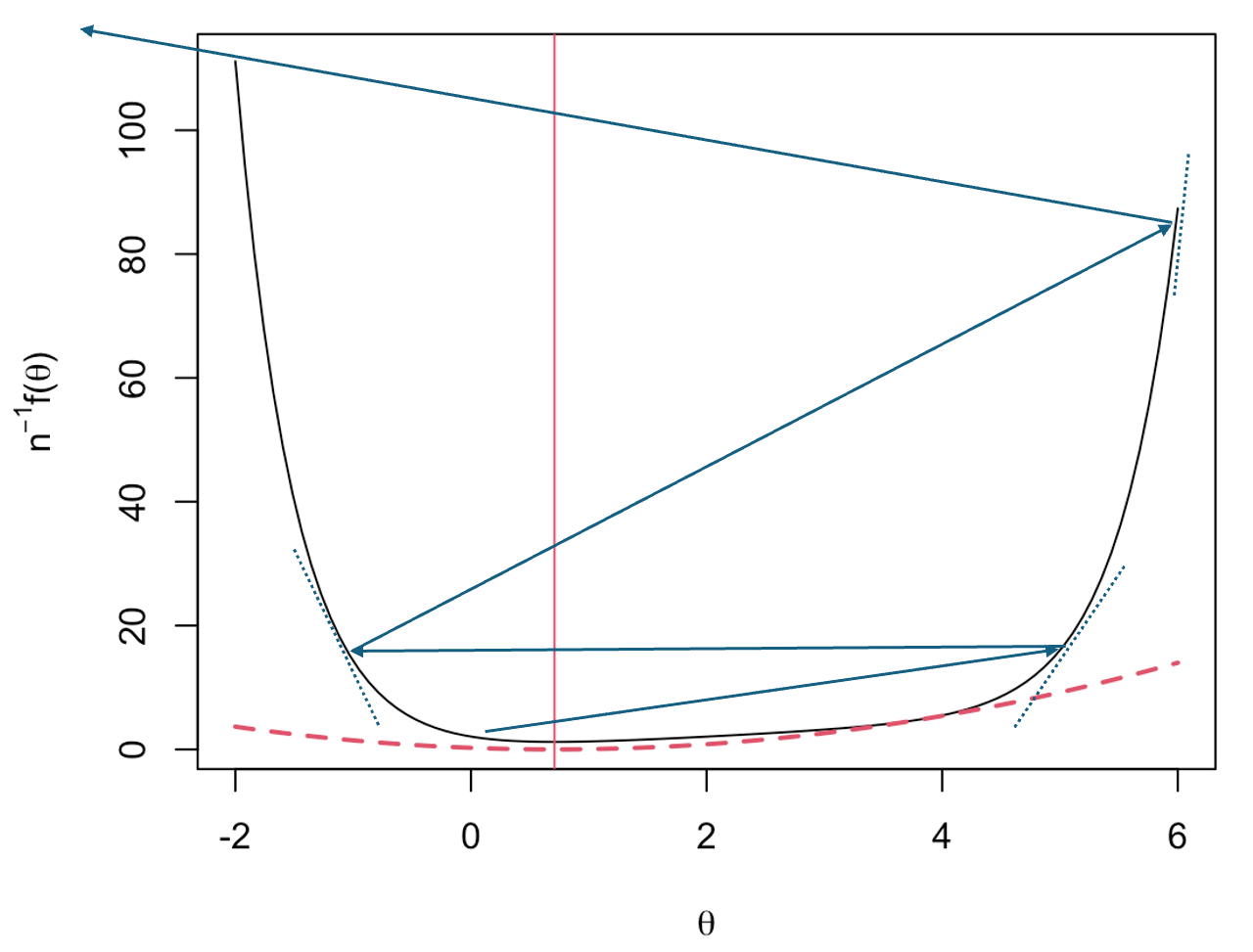}
    \caption{Schematic demonstration of numerical instabilities in Langevin sampling with $f'(\theta)$ \eqref{eq:grad_var}. 
    Due to the super-linear growth of $|f'(\theta)|$, iterates begin oscillating between increasingly farther two ends of the density.
    {\color{red} Red (dashed)} plots a quadratic function centered around the mode of $e^{-f(\theta)}$, 
    for which ULA is provably ergodic for a small enough $\epsilon$ and exact as $\epsilon\to0$.}
    \label{fig:fig2}
\end{figure}

The re-parameterization $\theta = \log\sigma$ appears innocent, as it is used by many probabilistic programming languages (e.g., Stan \cite{stanmanual}).
Yet it turns out that, based on the arguments of Section 3.2 \citep{roberts1996exponential}, the Markov chain of ULA or SGLD iterates is transient. 
In other words, \emph{algorithms \eqref{eq:ula}--\eqref{eq:sgld} do not sample from any distribution asymptotically, because they do not have a stationary distribution.} 
Furthermore, using the calculations of Lemma 6.3 \citep{mattingly2002ergodicity}, one can derive an exact expression for the probability that $|\theta_{k+1}|\geq \epsilon^{-1/2}2^k$ for every $k\geq 1$, which is never zero for any $\theta_0$ and any $\epsilon$, to arrive at the conclusion: 
\emph{when run long enough, Markov chains blow up with positive probability.} 
As empirical evidence, we show the iterates of the SGLD algorithm \eqref{eq:sgld} with $n = 1,000,\;S=5$ and $\epsilon = S/n^{1.4}$ in Figure \ref{fig:fig1}. 
The dataset $\{y_1,\ldots,y_n\}$ is simulated from a mean-zero normal distribution with true, unknown $\sigma = 2$. 
The Markov chain is initialized within a $1/\sqrt{n}$-neighborhood of $\log(2)$;
for a very long time, the iterates appear to be a good approximation of the posterior concentrated around $\log(2)$. 
However, at $k=1,021,459 \approx n^2$, the iterates reaches an unrealistically large value of $\theta$, 
after which the derivative and the updates become numerically unstable.
Although this instability is theoretically observable for ULA, our empirical experiments suggest that the additional noise of data subsampling in SGLD exacerbates the issue,
by accelerating the time at which the chain becomes numerically unstable.

While the theoretical reasoning of the previous citations are perfectly sound and not difficult to follow, the instability is surprising from a practical standpoint.
The key is that without a Metropolis-Hastings step, there is no automatic guarantee that the iterates of ULA or SGLD will converge. 
The stability of the Markov chain then critically depends on whether $f'$ is globally Lipschitz, which is often stated as a ``smoothness'' regularity condition in the theoretical analyses of ULA 
\cite{durmus2017nonasymptotic,durmus2019high,vempala2019rapid}.
However, as Figure \ref{fig:fig2} demonstrates, \eqref{eq:grad_var} is very far from being globally Lipschitz in $\theta$.

\section{Langevin Sampling and GLMMs}
\label{sec:model_desc}

The simulated experiment from the previous Section demonstrates that SGMCMC methods for unconstrained parameters are inadequate for use in constrained sampling that naturally arises in Bayesian inference.
In this Section, we will describe a scalable stochastic sampling algorithm based on a \emph{mirror} Langevin dynamics.
While the method is a generic constrained sampling algorithm, we will then specify the method to the problem of variance sampling in GLMMs.

\subsection{Mirror Langevin Dynamics}

Suppose $\theta\in\R^d$ has a distribution $\propto e^{-f(\theta)}$ supported on a convex subset $\Theta$ with boundary $\partial\Theta$. 
One first chooses a \emph{barrier function}, $\phi:\Theta\to\R$, such that $\phi(\theta)\to\infty$ whenever $\theta\to\theta_b\in\partial\Theta$.
We refer to the gradient $\nabla\phi$ as the \emph{mirror map}, and the image $\Theta^* = \nabla\phi(\Theta)$ as the \emph{dual space}.
Before proceeding, let us quickly describe the minimal requirements on $\phi$.

\begin{assumption}
    \label{assump1}
    $\phi : \Theta\to\R$ is twice continuously differentiable, strictly convex, and $\|\nabla^k\phi(\theta)\|\to\infty,\; k = 1,2$ as $\theta\to\theta_b$ for every $\theta_b\in\partial\Theta$.
\end{assumption}

Denote by $\phi^*(\vartheta) := \sup_{\theta\in\Theta}\langle\theta,\vartheta\rangle_{\R^d} - \phi(\theta)$ the convex conjugate of $\phi$. 
Under Assumption \ref{assump1}, this makes perfect sense and, furthermore, we have identities $\nabla\phi^* = (\nabla\phi)^{-1}$ and $\{\nabla^2\phi(\theta)\}^{-1} = (\nabla^2\phi^*\circ\nabla\phi)(\theta)$ for every $\theta\in\Theta$.
For brevity, we will write from hereon $V := \nabla f\circ\nabla\phi^*$ and $A:= \nabla^2\phi\circ\nabla\phi^*$.

Simulating mirror Langevin dynamics on the dual space $\Theta^*$ aims at discretizing the following SDE for a $d$-dimensional Brownian motion $B_t$:
\begin{equation}
    \label{eq:mld}
        d\vartheta_t = -V(\vartheta_t) + \sqrt{2A(\vartheta_t)}dB_t.
\end{equation}
The following \emph{mirror Langevin algorithm} \cite{li2022mirror} discretizes the SDE with a constant step size $\epsilon > 0$ and each $Z_k$ an independent ${\rm Normal}_d(0,I_d)$ random variable:
\begin{equation}
    \label{eq:mla}
    \vartheta_{k+1} = 
    \vartheta_k - 
    \epsilon V(\vartheta_k) + \sqrt{2\epsilon A(\vartheta_k)} Z_k.
\end{equation}
The $W_2$-distance between the chain above and the targeted posterior was shown to be $O(\sqrt{d\epsilon})$ \cite{li2022mirror}.
As we will mention later, this analysis is not immediately applicable to our work, since the dependence on $n$ will be hidden in the constants of the estimate.
An alternative discretization was also studied \cite{ahn2021efficient} using a different metric structure, results which we do not use in this work.

Our model will describe the $n$ terms corresponding to likelihoods of independent groups
and an extra term corresponding to a log-prior density,
so it is natural to assume the form
$\nabla f(\theta) = \sum_{i=0}^{n} \nabla f_i(\theta)$.
When $n$ is large, a natural data subsampling analogue to the algorithm \eqref{eq:mla} is to randomly draw $S>0$ indices with replacement and computing a stochastic gradient estimator. 
Denoting by $\mathcal{S}_k$ a minibatch of random indices independently drawn at $k$-th iteration, we can study a new algorithm, writing $V_i:=f_i\circ\nabla\phi^*$:
\begin{equation}
    \label{eq:smld}
    \vartheta_{k+1} = 
    \vartheta_k - \epsilon
    \left\{\frac{n}{S}\sum_{i\in\mathcal{S}_k} V_i(\vartheta_k) + V_0(\vartheta_k)\right\}
    + \sqrt{2\epsilon A(\vartheta_k)} Z_k.
\end{equation}
We refer to this algorithm as the {\it stochastic mirror Langevin dynamics (SMLD)} algorithm.
When $\Theta = \R^d$ and $\phi(\theta) = \|\theta\|^2/2$, the map $\nabla\phi$ is identity, so the underlying SDE \eqref{eq:mld} and its discretizations \eqref{eq:mla}--\eqref{eq:smld} reduce to their unconstrained analogues in \eqref{eq:sde}--\eqref{eq:sgld}. 
Hence, SMLD algorithm generalizes the usual SGLD algorithm.
To our knowledge, unlike SGLD, the stochastic subsampling analogue of MLD has not been studied.
The situation contrasts with the current literature in optimization, where mirror descent algorithm with stochastic gradient has been studied in the context of Markov decision processes \cite{jin2020efficiently} and policy optimization \cite{yang2022policy}.

\subsection{Generalized Linear Mixed Models}
\label{sec:glmm}

In a GLMM, each group $i\in[n]$ has $n_i\geq 1$ repeated measures, denoted by $y_{ij}$ for $j \in[n_i]$. 
Let $x_{ij}\in\R^p$ denote covariates and $z_{ij}\in\R^q$ denote covariates assumed to have effects on outcomes varying in group assignment.
The elements of the vector $y_i = (y_{i1},\ldots,y_{in_i})^\top$ are modeled as conditionally independent random variables from an exponential family distribution of density
$
p(y_{ij} | \eta_{ij}, \psi) = 
\exp\left[a(\psi)^{-1}\left\{y_{ij}
h(\eta_{ij}) - b(\eta_{ij})\right\} + c(y_{ij}, \psi)\right],
$
where $\eta_{ij}$ is the model's natural parameter, a function of the $ij$-th covariate, $\psi>0$ is a dispersion parameter,
and $a(\cdot) > 0$ and $b(\cdot),c(\cdot)$ are known, often smooth functions.
In particular, $b(\eta_{ij})$ relates to the mean model $\mu_{ij} := \E[y_{ij}|\eta_{ij}]$ by the relation $\mu = b'(\eta)/h'(\eta)$.
We choose a simple linear parameterization for the natural parameter: $\eta_{ij} := x_{ij}^\top\beta+ z_{ij}^\top\gamma_i$.
The parameter $\beta\in\R^p$ is a vector of population-level regression coefficients (``fixed effects'') and $\gamma_i\in\R^q,\; i \in[n]$ is each a vector of $i$-th group-specific deviations in the regression coefficients (``random effects''). 
Group-specific parameters are often assumed to have a ${\rm Normal}_q(0,\Sigma)$ distribution with unknown covariance $\Sigma$.
We will mostly focus on the Gaussian random effects model due to its widespread use and parameterize it using $\Omega = \Sigma^{-1}$, the precision matrix.
See Table \ref{tab1:exp} in the Appendix for the most commonly used models, to some of which we refer in this work.

We seek to approximate the marginal posterior distribution of population-level parameters $\theta := (\beta,\Omega,\psi)$, given a prior distribution $\pi_0$ for $\theta$ with a continuous density. We write $f_0 := -\log\pi_0$. For simplicity, let us exclude $\psi$ by setting $a = 1$. Then $\theta = (\beta,\Omega)$, where $\beta$ is an unconstrained real vector and $\Omega$ is $q\times q$ positive definite matrix. The cone of positive definite matrices is identifiable to a convex subset of $\R^{q(q+1)/2}$.
A natural barrier function which satisfies Assumption \ref{assump1} with $\alpha=1$ is $\phi^\Omega(\Omega) := -\log\det(\Omega)$. 
Defining $\phi(\beta,\Omega) = \|\beta\|^2/2 -\log\det(\Omega)$,
we can straightforwardly apply the SMLD update \eqref{eq:smld} to this GLMM with $\theta = (\beta,\Omega)\in\R^{p+q(q+1)/2}$ as follows:
\begin{align*}
    &\begin{pmatrix}
        \beta_{k+1} \\
        -\Omega_{k+1}^{-1} 
    \end{pmatrix}
    = \begin{pmatrix}
        \beta_k \\
        -\Omega_k^{-1} 
    \end{pmatrix}
    -\epsilon\frac{n}{S}\sum_{i\in\mathcal{S}_k}\begin{pmatrix}
    \nabla_\beta f_i(\beta_k)\\
    \nabla_\Omega f_i(\Omega_k)
    \end{pmatrix}\\
    &\quad-\epsilon\begin{pmatrix}
        \nabla_\beta f_0(\beta)\\
        \nabla_\Omega f_0(\Omega)
    \end{pmatrix}
    +\sqrt{2\epsilon}\begin{pmatrix}
        Z_k\\
        \sqrt{\Omega_k^{-1}}W_k\sqrt{\Omega_k^{-1}}^\top
    \end{pmatrix}.
\end{align*}
Here, each $Z_k$ is a $p$-dimensional standard normal random vector and each $W_k$ is a $q\times q$-dimensional symmetric random matrix whose diagonal and off-diagonal entries are standard normal.
The summand $f_i$ corresponds to the (negative) \emph{marginal log-likelihood} of the $i$-th group:
\begin{equation*}
    f_i(\theta) 
    = -\log\int_{\R^q}
    \underbrace{\left\{\prod_{j=1}^{n_i}p(y_{ij}|\gamma_i,\theta)\right\}
    p(\gamma_i|\theta)}_{=p(y_i,\gamma_i|\theta)}~d\gamma_i.
\end{equation*}
In principle, we have completed the prescription of the use of SMLD for GLMM inference. However, many algorithmic details have been left out.
One immediate issue is that the multivariate integral in the above display is generally intractable.
In the next Section, we discuss the issue of computing the stochastic gradient estimator and correctly calibrating the posterior sample variance.

\section{Application of SMLD to GLMM Inference}
\label{sec:mirror}

\subsection{Stochastic Gradient Estimation}

To implement the SMLD update \eqref{eq:smld}, it is not necessary to evaluate pointwise the intractable integral $f_i$.
We do need to evaluate pointwise the \emph{gradient} of this integral, 
but this is a task much more amenable to a sampling-based approximation.
Concretely, by Fisher's identity \cite{fisher1925theory}, exchanging the derivative and the integral yields
    $\nabla f_i(\theta) = 
    -\E_{p(\gamma_i|y_i,\theta)}[\nabla_\theta\log p(y_i,\gamma_i|\theta)]$,
where the conditional distribution $p(\gamma_i|y_i,\theta)$ is given by
\begin{equation}    \label{eq:full_conditional}
    p(\gamma_i|y_i,\theta) = 
    \frac{p(y_i|\gamma_i,\theta)
    p(\gamma_i|\theta)}{\int_{\R^q}p(y_i|\xi,\theta)
    p(\xi|\theta)~d\xi}.
\end{equation}
For now, assume we can draw \emph{independent} Monte Carlo samples from the conditional distribution $p(\gamma_i|y_i,\theta)$.
Then, for each $i$, an unbiased Monte Carlo estimator of $g_i$ using $R > 0$ samples is given by
\begin{equation}
    \label{eq:grad_i_est}
    \widehat{\nabla f}_i^{(R)}(\theta) := -\frac{1}{R}\sum_{r=1}^{R}\nabla\log p(y_i,\gamma_{ir}|\theta),\;
    \gamma_{ir}\stackrel{iid}{\sim} 
    p(\gamma_i|y_i,\theta).
\end{equation}
This is a natural stochastic gradient estimator for $\nabla f_i(\theta)$, so we may modify the previous SMLD update into
\begin{equation}
    \label{eq:smld_glmm}
    \vartheta_{k+1} = 
    \vartheta_k - \epsilon
    \left\{\frac{n}{S}\sum_{i\in\mathcal{S}_k}
    \widehat{V}_i^{(R)}(\vartheta_k) + V_0(\vartheta_k)\right\} 
    +\sqrt{2\epsilon A(\vartheta_k)}Z_k,
\end{equation}
where we wrote $\widehat{V}_i^{(R)} := \widehat{f}^{(R)}_i\circ\nabla\phi^*$.

For a later development, we will also derive the covariance matrix of the stochastic gradient estimator. 
Simple computations show that the covariance of stochastic gradient \eqref{eq:grad_i_est} is given by
\begin{equation}
    \label{eq:cov_grad_est}
    \Psi_i^{(R)}(\theta) := \frac{1}{R}
    \mathbb{V}_{p(\gamma_i|y_i,\theta)}
    [\nabla_\theta\log p(y_i,
    \gamma_i|\theta)].
\end{equation}
Since $\widehat{\nabla f}_i^{(R)}(\theta)$ is unbiased with respect to $f_i(\theta)$ at each $\theta$,
it is clear that for a random minibatch of indices $\mathcal{S}_k$ drawn with replacement,
$S^{-1}\sum_{i\in\mathcal{S}_k}\widehat{\nabla f}_i^{(R)}$ is unbiased with respect to $n^{-1}\nabla f$.
The total covariance of $S^{-1}\sum_{i\in\mathcal{S}_k}\widehat{\nabla f}_i^{(R)}(\theta)$ is given by $\Psi^{(R)}(\theta)/S$, for
\begin{equation}
    \label{eq:cov_grad_full}
    \Psi^{(R)}(\theta) := \frac{1}{n}\sum_{i=1}^{n}
    \left[\left\{\nabla f_i(\theta) - \frac{1}{n}\nabla f(\theta)\right\}^{\otimes 2}
    + \Psi_i^{(R)}(\theta)\right].
\end{equation}
Writing by $\widehat{\Psi}_i^{R}(\theta)$ the Monte Carlo-based unbiased estimator of \eqref{eq:cov_grad_est}
and $\widehat{\nabla f}^{(R)} := \sum_{i=1}^{n}\widehat{\nabla f}^{(R)}_i$,
the following gives an unbiased estimator of $\Psi^{(R)}(\theta)$:
\begin{equation}
    \label{eq:cov_grad_full_est}
    \widehat{\Psi}^{(R)}(\theta) = \frac{1}{n}\sum_{i=1}^{n}\left[
    \left\{\widehat{\nabla f}_i^{(R)}(\theta) - \frac{\widehat{\nabla f}^{(R)}(\theta)}{n}\right\}^{\otimes2} + \frac{\widehat{\Psi}_i^{(R)}(\theta)}{n}\right].
\end{equation}
Computations of \eqref{eq:cov_grad_est} and \eqref{eq:cov_grad_full} are elementary; see Section \ref{sec:proof_misc} for the unbiasedness proof of \eqref{eq:cov_grad_full_est}.

So far, we have assumed the ability to draw independent Monte Carlo samples from the random effects conditional distribution.
We will maintain this assumption for the theoretical analysis to follow in this Section.
However, the normalizing constant in the distribution \eqref{eq:full_conditional} is intractable for most GLMMs, so independent Monte Carlo samples are mostly unattainable.
In such cases, we will use $R$ \emph{dependent} samples $\{\gamma_{ir}\}_{r=1}^{R}$ obtained from an MCMC, ergodic with respsect to \eqref{eq:full_conditional}, 
in computing the stochastic gradient \eqref{eq:grad_i_est}.
This poses a technical complication in the theory, as the stochastic gradient \eqref{eq:grad_i_est} is then no longer unbiased.
The difficulties of a fully rigorous analysis, along with some special cases for which an unbiased Monte Carlo stochastic gradient is computable, are sketched out in Section 3.3 \citep{berchuck2026scalablebayesianinferencegeneralized}.
% of the Supplementary Materials.
Nevertheless, empirical results in Sections \ref{sec:sim_study} and \ref{sec:real_study} suggest our method can approximate the posterior well for a modestly large $R$ and the choice of MCMC algorithm with good mixing properties,
and thus the gap between theory and practice need not be so great.

\subsection{Error Analysis}

We now turn to an error analysis of the SMLD algorithm.
First, we state the regularity conditions on the mirror map and the targeted posterior density.

\begin{assumption}
    [Modified Self-Concordance]
    \label{assump:selfconcordance}
    The mapping $A = \nabla^2\phi\circ\nabla\phi^*$ satisfies
        $\left\|\sqrt{A(\vartheta)}-\sqrt{A(\vartheta')}\right\|_{HS}
        \leq \sqrt{\alpha}
        \|\vartheta - \vartheta'\|$
    for every $\vartheta,\vartheta'\in\Theta^*$ and some $\alpha > 0$.
\end{assumption}

\begin{assumption}
    [Relative Smoothness]
    \label{assump:smoothness}
    For all $i\in \{0\}\cup[n]$, $V_i = \nabla f_i\circ\nabla\phi^*$ satisfies
        $\|V_i(\vartheta) - V_i(\vartheta')\| \leq L\|\vartheta - \vartheta'\|$
    for every $\vartheta,\vartheta'\in\Theta^*$ and some $L>0$.
\end{assumption}

\begin{assumption}
    [Relative Strong Convexity]
    \label{assump:strong_logconcavity}
    For all $i\in [n]$, $V_i = \nabla f_i\circ\nabla\phi^*$ satisfies
        $\langle V_i(\vartheta) - V_i(\vartheta'), \vartheta - \vartheta'\rangle
        \geq m\|\vartheta - \vartheta'\|^2$
    for every $\vartheta,\vartheta'\in\Theta^*$ and some $m > 0$.
    Furthermore, $V_0$ is convex.
\end{assumption}
A quick consequence of Assumption \ref{assump:smoothness} is that $V$ satisfies $\|V(\vartheta) - V(\vartheta')\|\leq(n+1)L\|\vartheta - \vartheta'\|$ for every $\vartheta,\vartheta'\in\Theta^*$.
Similarly, by Assumption \ref{assump:strong_logconcavity}, $V$ satisfies $\langle V(\vartheta_1) - V(\vartheta_2),\vartheta_1 - \vartheta_2\rangle\geq nm\|\vartheta_1 - \vartheta_2\|^2$.
Finally, continuity of $\nabla f,\nabla\phi^*$ and  \ref{assump:strong_logconcavity} guarantees the unique existence of a global minimizer of $V$. 
This is the MLE of the model, implicitly dependent on $n$, which we hereby denote by $\vartheta^*$.

The following result quantifies the approximation error of SMLD with respect to the target posterior.

\begin{theorem}
    \label{thm:w2_err}
    Let $\pi\propto e^{-f}$ be the targeted posterior distribution $\propto e^{-f}$ on $\Theta$,
    with $f=\sum_{i=0}^{n+1}f_i$ and $\phi$ meeting Assumptions \ref{assump:selfconcordance}--\ref{assump:strong_logconcavity}.
    Assume that $\vartheta^*$, the unique minimizer of $V$,
    belongs to the interior of a bounded subset of $\Theta^*$ for all $n\geq 1$, and that $\E_\pi\|\vartheta\|^2 = O(1)$.
    Let $\rho_k$ be the marginal distribution of $\vartheta_k$ in SMLD algorithm \eqref{eq:smld}, with $\rho_0\in\mathcal{P}^2(\R^d)$.
    For $\epsilon = O(n^{-(1+\delta)}),\;\delta\in(0,1)$ and for all sufficiently large $n$, we have
    \begin{equation*}
    W_2(\rho_k,\nabla\phi_\#\pi) \leq O\left(e^{-knm\epsilon/4}W_2(\rho_0,\nabla\phi_\#\pi) + \sqrt{\epsilon n}\right).
    \end{equation*}
\end{theorem}
See Section \ref{sec:proof1} of the Appendix for a proof.

\begin{remark}
    Theorem 1 \cite{li2022mirror} states a Wasserstein-distance error bound of the above type for the algorithm \eqref{eq:mla}.
    However, this result in its literal form is not useful in our context, because its dependence on the $O(n)$ smoothness and strong convexity constants of $f$ is very poor.
    The proof techniques we used are not novel and should be familiar to readers of previous literature \cite{durmus2017nonasymptotic,durmus2019high,dalalyan2019user}.
    Since we are interested in regimes where $n\gg d$, the proof does not attempt to optimize dependence on $d$.
\end{remark}

\subsection{Sample Variance Post-Processing}

Theorem \ref{thm:w2_err} states that when $\epsilon \asymp n^{-(1+\delta)},\; \delta>0$ and the chain is run for length $k\gg n^{1+\delta}$,
the squared distance between the SMLD Markov chain's law and the underlying posterior decays like $O(n^{-\delta})$.
On the other hand, Laplace approximation suggests that the true posterior distribution variance must be like $O(n^{-1})$, at least if all MLEs remain in the interior of $\Theta$ for large $n$.
Inspection of the proof shows that the additional variance of the stochastic gradient \eqref{eq:grad_i_est} dominates the discretization error of the no-subsampling algorithm \eqref{eq:mla} and causes an order-wise larger error.
We will propose here a non-intrusive post-processing algorithm of the SMLD samples.
Assuming the Laplace approximation estimates the true posterior variance with $O(n^{-1})$ error, our method yields an asymptotically, order-wise correct estimate of the underlying posterior variance, while avoiding the practical difficulties of designing control variates and initialization for GLMMs, highlighted in Section 4.2 \citep{berchuck2026scalablebayesianinferencegeneralized}.
% of the Supplementary Materials.

Assuming again the stochastic gradient estimator \eqref{eq:grad_i_est} is unbiased, let us re-write the SMLD update as
\begin{equation}
    \label{eq:smld_re}
    \vartheta_{k+1} = 
    \vartheta_k - 
    \epsilon V(\vartheta_k) +
    \sqrt{2\epsilon}
    \left\{\sqrt{\frac{\epsilon}{2}}\xi(\vartheta_k,\mathcal{S}_k) + \sqrt{A(\vartheta_k)}Z_k\right\},
\end{equation}
where the random vector $\xi(\vartheta_k,\mathcal{S}_k)$ is defined so that the above display is equal to \eqref{eq:smld_glmm}.
Based on the moment computations \eqref{eq:cov_grad_est}--\eqref{eq:cov_grad_full},
this random vector has a covariance matrix $(\epsilon n^2)/(2S)\Psi^{(R)}(\nabla\phi^*(\vartheta_k))$. For each $\vartheta$, we define a new covariance matrix
\begin{equation}
    \label{eq:grad_noise_cov}
    \Gamma^{(R)}(\vartheta) := \frac{\epsilon n^2}{2S}
    \Psi^{(R)}(\nabla\phi^*(\vartheta)) + 
    A(\vartheta);
\end{equation}
which is the ``total covariance'' of the noise component at each SMLD iteration.
Importantly, for any fixed value of $\vartheta$ and thus $\theta = \nabla\phi^*(\vartheta)$, the matrix above can be approximated using a full scan of the $n$-sized dataset and $R$ Monte Carlo/MCMC samples using the formula \eqref{eq:cov_grad_full_est}.
A naive second-order Taylor expansion around the transformation of MLE, $\vartheta^* = \nabla\phi(\theta^*)$, 
along with simplifying the distribution of the random vector $\xi'(\vartheta_k;\mathcal{S}_k)$ to be also Gaussian,
leads us to believe that the update \eqref{eq:smld_re} must approximate the following SDE on the dual space:
\begin{equation*}
    d(\vartheta_t - \vartheta^*) = 
    -HJ^{-1}(\vartheta_t-\vartheta^*)dt + 
    \sqrt{2\Gamma(\vartheta^*)}dB_t,
\end{equation*}
where $H = \nabla^2 f(\theta^*)$ and $J = \nabla^2\phi(\theta^*)$.
The invariant distribution of this SDE is ${\rm Normal}_d(\vartheta^*,V)$ whose covariance solves the Lyapunov equation
\begin{equation}
    \label{eq:lyapunov}
    XJ^{-1}V + VJ^{-1}X = 2\Gamma^{(R)}(\theta^*)
\end{equation}
with $X=H$.
A more detailed, step-by-step reasoning of this heuristic in the context of unconstrained sampling is given in Section 3.2 \citep{berchuck2026scalablebayesianinferencegeneralized}.
% of the Supplementary Materials.

On the other hand, consider a Gaussian approximation to the true posterior in the large-$n$ regime.
By a Laplace approximation argument, we expect the true posterior distribution of $\theta$ to look like ${\rm Normal}_d(\theta^*,H^{-1})$ when $n$ is large.
By the delta method, we expect the distribution of $\nabla\phi(\theta)$ to look like ${\rm Normal}_d(\nabla\phi(\theta^*), JH^{-1}J)$.
The upshot is that given the quantities $\theta^*,\Gamma(\theta^*)$ and $V$, 
we can solve the equation in unknown $X$ and obtain $H^{-1}$, which will in turn approximate the posterior covariance for large $n$.

The following result demonstrates that if we take the Laplace approximation to be accurate enough,
then the heuristic reasoning behind \eqref{eq:lyapunov} is sound in the large-$n$ regime ``up to a vanishing error relative to $\Gamma^{(R)}(\theta^*)$,''
providing us a way to estimate the curvature $H$.
For this result, we assume a stronger smoothness condition than Assumption \ref{assump:smoothness}.

\begin{assumption}
    \label{assump:smoothness_str}
    For all $i\in \{0\}\cup[n]$, $V_i$ is twice continuously differentiable and $\|D^2V_i\|\leq L$, where the norm is the operator norm for 3-linear forms in $\Theta^*$.
\end{assumption}

\begin{theorem}
    \label{thm:lyapunov_error}
    Suppose that $\vartheta^*$, the unique minimizer of $V$,
    belongs to the interior of a bounded subset of $\Theta^*$ for all $n\geq 1$.
    Under Assumption \ref{assump:smoothness_str},
    for $\epsilon = O(n^{-(1+\delta)}),\;\delta\in(0,1)$ and for all sufficiently large $n$,
    the covariance $V$ of the invariant distribution of SMLD \eqref{eq:smld_glmm} satisfies
    \begin{equation*}
        HJ^{-1}V + VJ^{-1}H = 2\Gamma^{(R)}(\theta^*)\{1+O(n^{-\delta/2})\}
    \end{equation*}
    in the sense of operator norms for $d\times d$ matrices.
\end{theorem}
See Section \ref{sec:proof2} of the Appendix for a proof.

\begin{remark}
    We do not address here the issue of whether the error between $H^{-1}$, the inverse curvature around the MLE, and the true posterior variance of $\theta$ is indeed $O(n^{-1})$.
    In the context of simpler i.i.d. likelihood models, finite sample bounds in terms of $n$ and $d$ were provided by \cite{kasprzak2025good}.
    Under Assumptions \ref{assump:smoothness} and \ref{assump:strong_logconcavity}, we can use simpler results along the lines of Proposition 5 \cite{brosse2018promises}.
    Despite the folklore status of GLMMs in statistics,
    it seems difficult to find a quantitative statement on the asymptotic normality of likelihood-based estimators, as mentioned by the unpublished talk \cite{Johnstone24vid}.
\end{remark}

To summarize, solving the Lyapunov equation \eqref{eq:lyapunov} in $X$ permits us to approximate the posterior covariance of $\theta$ or $\nabla\phi(\theta)$.
Since $\vartheta^*$ and $V$ are not exactly known, we approximate them by the mean and covariance matrix of the SMLD samples, respectively;
we denote these sample-based approximations by $\widehat{\vartheta^*}$ and $\widehat{V}$, respectively,
while $\Gamma^{(R)}(\theta^*)$ also permits a natural esitmator as mentioned \eqref{eq:cov_grad_full_est}.
Denoting the solution of the resulting equation by $\widehat{H}$, we can re-scale the posterior samples $\vartheta_1,\vartheta_2,\ldots$ obtained from the SMLD updates \eqref{eq:smld_glmm} to calibrate the uncertainty estimates:
\begin{equation}
    \label{eq:rescale}
    \vartheta_k^{\rm corr}\gets \widehat{\vartheta^*} + \widehat{J}\widehat{H}^{-1/2}\widehat{V}^{-1/2}(\vartheta_k - \widehat{\vartheta^*}).
\end{equation}
All these steps, together with the implementation of the SMLD \eqref{eq:smld} in the context of GLMMs, are summarized in Algorithm \ref{alg} in the Appendix.

\begin{figure*}[!t]
    \centering
    \includegraphics[width=0.75\linewidth]{Simulation2_Meanplot.png}
    \caption{Posterior mean estimates presented across runtime (hours) for the Bernoulli GLMM with simulated datasets (Section \ref{sec:sim_study2}). Columns and rows indicate parameter and sample size (n), respectively; for each panel, posterior means were averaged over 30 different datasets.
    Thick dashed line indicates the true parameter values used to simulate the data.}
    \label{fig:figsim2_mean}
\end{figure*}

\begin{figure*}[!h]
    \centering
    \includegraphics[width=0.75\linewidth]{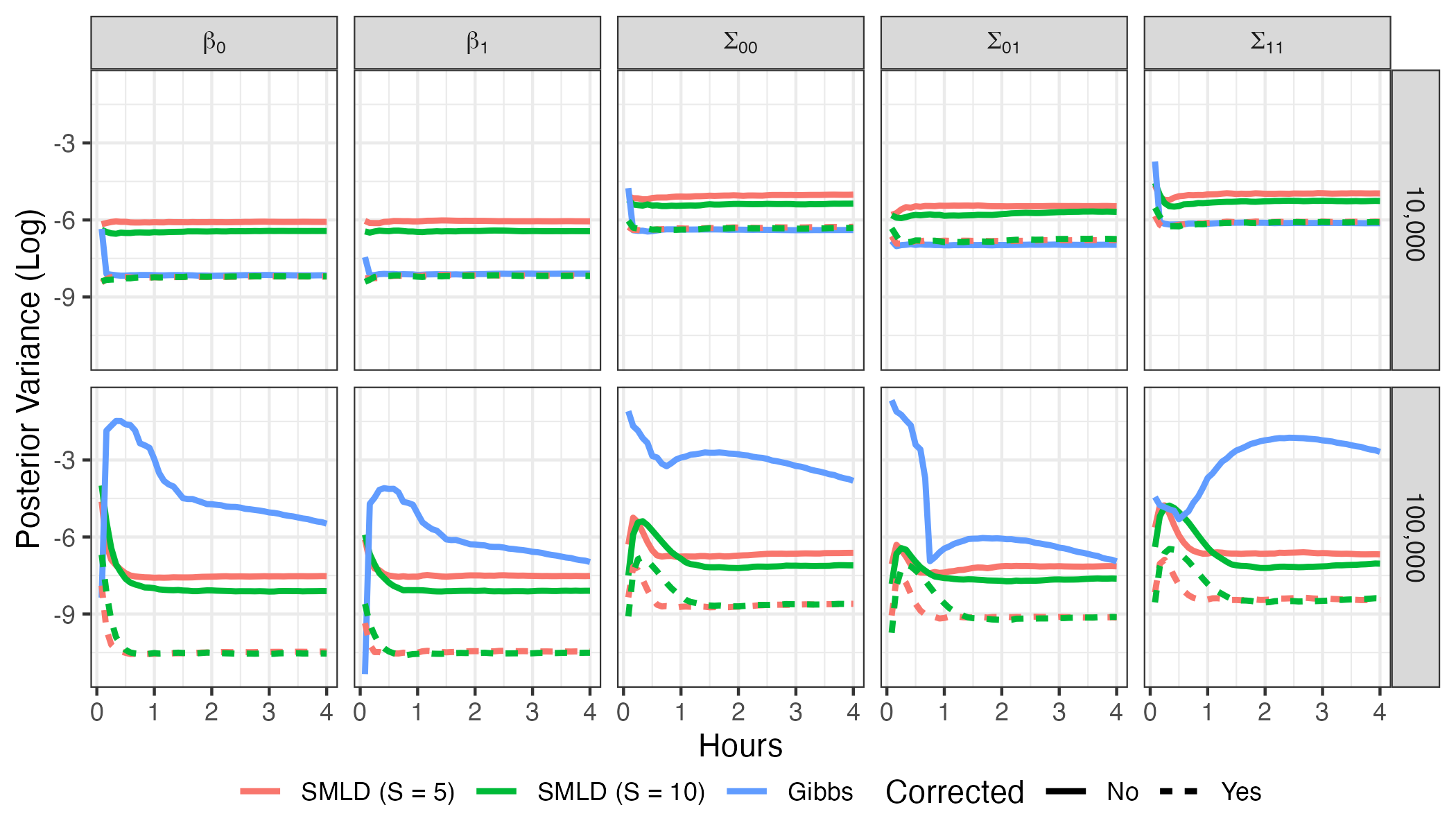}
    \caption{Log posterior variance estimates presented across runtime (hours) for the Bernoulli GLMM with simulated datasets (Section \ref{sec:sim_study2}). Columns and rows indicate parameter and sample size (n), respectively; for each panel, posterior variances were averaged over 30 different datasets.}
    \label{fig:figsim2_var}
\end{figure*}

\section{Simulated Experiments}
\label{sec:sim_study}

We present simulated studies that validate the empirical performance of the corrected SMLD algorithm. 
% Codes used to perform simulated data studies in Sections \ref{sec:sim_study1} and \ref{sec:sim_study2} are included in the Supplementary Materials.

\subsection{Toy Model Demonstration}
\label{sec:sim_study1}

We first validate the correctness of the mean and covariance of the corrected SMLD samples based on a toy model where an exact posterior is known.
The distribution to be sampled from is a posterior given a multivariate normal likelihood for data $y_1,\ldots,y_n\in\R^2$, known to be mean zero and with unknown covariance matrix $\Sigma$,
and an ``uninformative'' ${\rm Wishart}(2,I_2)$ prior on the precision matrix $\Sigma^{-1}$.
50 different datasets comprising random vectors $y_1,\ldots,y_n\in\R^2$, each of size $n=5,000$, were generated from a multivariate normal distribution with $\Sigma = 
\begin{bmatrix}
    1.5 & .25 \\ .25 & 1.5
\end{bmatrix}$.
Setting the minibatch size $S=5$, we run the SMLD algorithm \eqref{eq:smld} with step size $\epsilon = S/n^{1.3}$ for a total length of $\lceil20/\epsilon\rceil$ iterations, before correcting the vectorized covariance matrix samples as described in Algorithm \ref{alg}.
Table \ref{tab2:sim1} in the Appendix summarizes for each entry of $\Sigma$ the absolute error between true posterior moments and their SMLD sample-based approximations.
Our correction method by design does not change the SMLD sample mean, but it brings the sample covariance significantly closer to the true posterior covariance. 
The uncorrected SMLD sample-based variance estimates have a large bias mainly because the variance estimates themselves are too large compared with the true posterior variances.

\subsection{Simulated Data GLMMs}
\label{sec:sim_study2}

In this second simulation study, we consider the task of fitting a binary response GLMM on large simulated data with a logit link function (cf. Table \ref{tab1:exp}, Appendix).
We simulated 30 datasets for group sizes $n\in\{10^4,10^5\}$, each group having 10 equal observations and priors $\beta\sim{\rm Normal}(0,10^2I_2),\; \Sigma^{-1}\sim{\rm Wishart}(2,I_2)$.
We ran the SMLD algorithm, implemented in C++, with minibatch sizes $S\in\{5,10\}$ for 4 hours.
To validate the accuracy of the method, we compared both uncorrected and corrected samples against the true parameters used to generate the data and samples obtained from a full-batch Gibbs sampler based on a Polya-Gamma data augmentation scheme \cite{polson2013bayesian}, 
of which the true posterior is the invariant distribution.
The same data augmentation scheme was also used by the SMLD sampler in estimating the stochastic gradient \eqref{eq:grad_i_est} by drawing $R=1,000$ samples of $\gamma_i$s.
For each configuration of $n$ and $S$, we chose the step size to be $\epsilon=S/n^{1+\delta}$ with $\delta = (\delta_{min} + 1)/2$, where $\delta_{min} = \min_\delta\{S/n^{1+\delta} < 1/n\}$,
heuristically balancing the requirements of stable variance estimation and computational efficiency.
Figures \ref{fig:figsim2_var} and \ref{fig:figsim2_mean} 
% in the Appendix 
present the resulting posterior mean and variance estimates against run time.
Regardless of $n$ and $S$, SMLD sample-based posterior mean approximation quickly converges to a value close to the true values, but the variance approximation has an upward bias, which our correction method eliminates. 
When $n=10,000$, SMLD sample variance agrees with the variance of Gibbs sampler draws.
However, the two samplers disagree when $n=100,000$, due to the Gibbs sampler chain having produced too few samples to have reached convergence even after 4 hours, 
evidently seen from the fact that they do not produce a close enough mean to the true values.
Our study highlights that the corrected SMLD thus produces accurate enough posterior mean and variance estimates with a substantial computational gain over the conventional MCMC.

\begin{figure}[!t]
    \centering
    \includegraphics[width=0.9\columnwidth]{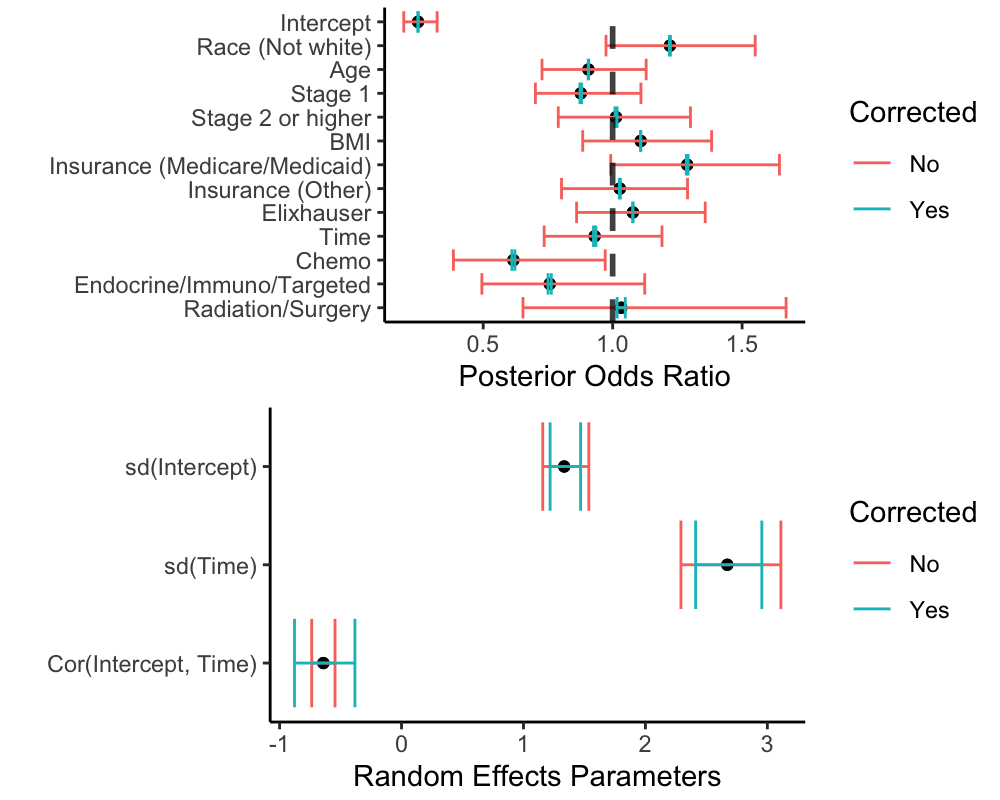}
    \caption{Posterior odds ratios of fixed effects and random effects covariance matrix parameters with 95\% credible intervals for Section \ref{sec:real_study}.}
    \label{fig:figdata}
\end{figure}

\section{Real-World Data Analysis}
\label{sec:real_study}

We now present results from fitting a binomial response GLMM on real-world biomedical dataset.
The data are collected from a study of breast cancer survivors in Duke Oncology clinic.
% [{\it a large academic cancer center}]. % anonymized
The outcome $y_{ij}\in\{0,1,\ldots,10\}$ of the $i$-th patient at $j$-th encounter is the reported pain score between 0 and 10, at each oncology encounter up to 1 year from initial diagnosis of cancer.
Patients who have only reported zero pain or have only one pain score available were excluded, leaving a dataset of $n=2,528$ patients and a total of $33,327$ integer-valued outcomes. 
We fit a GLMM with a $M=10$ binomial link function (cf. Table \ref{tab1:exp}, Appendix) that adjusts for covariates $x_{ij}\in\R^{13}$ including
each patient's race, age (years), BMI ($\text{kg/m}^2$), stage at initial diagnosis (baseline 0), insurance status (baseline commercial), and Elixhauser comorbidity index,
together with active treatment categories at each encounter and years relative to the initial diagnosis as a linear regressor.
The model included a random intercept and a random linear slope of time for modeling patient-specific increase and decrease in pain ($z_{ij}\in\R^2$).
Figure \ref{fig:figdata} plots the credible intervals for posterior fixed effects odds ratios (positive probability defined by a pain score above 5) credible and random effects covariance parameters.
For interpretability, intervals for the latter group of parameters are shown on marginal standard deviation and correlation scale.
The widths of the corrected credible intervals are significantly narrower for the fixed effects than for the random effects parameters and suggests all covariates have a significant cohort-level effect.
The widths narrower than $O(1/\sqrt{n})$ are caused by the fact that unlike in simulated studies, many patients have very dense trajectories (241 being the most data per patient and 57 patients with more than 50 pain scores).
Credible intervals of random intercept and random linear time slope, on the other hand, suggest a much larger between-patient variability than the significant cohort-level decrease of pain in time.
Interestingly, the credible interval for the correlation between random intercept and random linear slope in time becomes wider when corrected.
Ultimately, the dramatic changes in the credible intervals for the fixed effects support the importance of variance correction for correctly calibrating Bayesian $p$-values.

\section{Discussion}
\label{sec:discussion}

In this paper, we introduced a stochastic gradient MCMC algorithm suitable for constrained sampling in large datasets.
Specializing the general task to Bayesian inference, we have shown the importance of a tailored constrained sampling method for GLMMs and demonstrated how to implement the corrected SMLD algorithm for a calibrated posterior variance estimate.
Simulations demonstrate the accuracy and the clear computational gains of our proposed method for a popularly used binary observation GLMM with logit link function.

Our work has limitations and possible improvements.
First, while our justification of the use of MCMC was mainly empirical, a finer theoretical study, possibly very model-specific, is required to understand the precise effect of using a biased stochastic estimator.
In a similar vein, it will be also interesting to study the efficacy of alternatives to MCMC, such as quasi-Monte Carlo importance sampling \cite{he2023error} or expectation propagation \cite{hall2020fast}.
Second, we have not formally tackled the issue of a tightly optimized finite-sample bound, which can guide the choice of step size $\epsilon = O(n^{-(1+\delta)})$, which we have tuned mostly by trial and error.
Since the SMLD iterates \eqref{eq:smld} do not strictly enforce the parameter constraints,
a small enough choice of $\epsilon$ is very important for algorithmic stability,
while the user would also hope the targeted distribution also permits not too small an $\epsilon$ that will waste computational power.
The proof techniques we have adopted in this work may be too crude for this purpose and need improvements.
Finally, from a practical perspective, extensions of our algorithm towards incorporating momentum and fitting high-dimensional neural net models will form practically valuable contributions to Bayesian deep machine learning.

% Acknowledgements should only appear in the accepted version.
%\section*{Acknowledgements}

%%%%%%%%%%%%%%%%%%%%%%%%%%%%%%%%%%%%%%%%%%%
% DOES NOT COUNT INTO 8 PAGES FROM HEREON %
%%%%%%%%%%%%%%%%%%%%%%%%%%%%%%%%%%%%%%%%%%%

% \section*{Impact Statement}
% This paper introduces a scalable method for sampling problems in generalized linear mixed models. 
% While our work is primarily methodological and theoretical, its integration into diverse applications, including biomedical data analysis exemplified in this work, can lead to improved uncertainty quantification for large datasets and enhanced policy and decision-making. 
% We emphasize that care is needed for a practical deployment of the proposed method and that further studies are needed to address computational improvements and theoretically well-founded extensions of the proposed method to a broader class of models.

\bibliography{main}
\bibliographystyle{icml2026}

%%%%%%%%%%%%%%%%%%%%%%%%%%%%%%%%%%%%%%%%%%%%%%%%%%%%%%%%%%%%%%%%%%%%%%%%%%%%%%%
%%%%%%%%%%%%%%%%%%%%%%%%%%%%%%%%%%%%%%%%%%%%%%%%%%%%%%%%%%%%%%%%%%%%%%%%%%%%%%%
% APPENDIX
%%%%%%%%%%%%%%%%%%%%%%%%%%%%%%%%%%%%%%%%%%%%%%%%%%%%%%%%%%%%%%%%%%%%%%%%%%%%%%%
%%%%%%%%%%%%%%%%%%%%%%%%%%%%%%%%%%%%%%%%%%%%%%%%%%%%%%%%%%%%%%%%%%%%%%%%%%%%%%%
\newpage
\appendix
\onecolumn

\section{Tables and Figures}

\begin{table}[h!]
\centering
\caption{Some commonly used GLMM specifications. See Section \ref{sec:glmm} for notation ($\Phi$ is the standard normal CDF).}
\begin{tabular}{|l|c|c|c|c|c|} 
 \hline
 Model & Range of $y$ & $\E[y|\eta]$ & Cumulant $b(\cdot)$ & $h(\cdot)$ & $a(\cdot)$ \\ [0.5ex] 
 \hline\hline
 Gaussian (Linear) & $\R$ & $\eta$ & 
 $\frac{1}{2}\eta^2$ & Identity & Identity
 \\ 
 Logit: Binomial ($M\geq1$ trials) & $[M]$ & $\frac{Me^\eta}{1+e^\eta}$ & $M\log(1+e^\eta)$ & Identity & $a=1$ \\
 Probit: Bernoulli & $\{0,1\}$ & $\Phi(\eta)$ &
 $-\log\{1-\Phi(\eta)\}$ & 
 $h(\eta) = \log\frac{\Phi(\eta)}{1-\Phi(\eta)}$ & $a=1$\\
 Poisson & $\{0,1,2,\ldots,\}$ & $e^\eta$ & $e^\eta$ & Identity & $a=1$\\
 \hline
\end{tabular}
\label{tab1:exp}
\end{table}

\begin{table}[h!]
\centering
\caption{Table of results for simulated study in Section \ref{sec:sim_study1}.
Each row displays the 1st and 3rd quantiles from 50 different datasets.}
\begin{tabular}{|l|c|c|c|} 
 \hline
 & $\Sigma_{11}$ & $\Sigma_{12}$ & $\Sigma_{22}$ \\ [0.5ex] 
 \hline\hline
 True Values & 1.5 & 0.25 & 1.5 \\
 \hline
 Posterior Mean & [1.47, 1.51] & [0.23, 0.26] & [1.49, 1.52] \\
 \hline
 Absolute Error: Sample Mean & [6e-4, 2.4e-3] & [4e-4, 1.6e-3] & [9e-4, 3e-3]\\
 \hline
 \hline
 Posterior Variance & [9e-4, 9e-4] & [5e-4, 5e-4] & [9e-4, 9e-4] \\
 \hline
 Absolute Error: Uncorrected Sample Var & [0.09, 0.10] & [0.05, 0.05] & [0.10, 0.10] \\
 \hline
 Absolute Error: Corrected Sample Var & [1.4e-4, 1.6e-4] & [0.2e-4, 0.3e-4] & [1.4e-4, 1.6e-4] \\ [1ex] 
 \hline
\end{tabular}
\label{tab2:sim1}
\end{table}

\section{Algorithm Description}
\begin{algorithm}[!h]
\caption{SMLD with Covariance Correction for GLMMs}\label{alg}
\begin{algorithmic}
\STATE {\bfseries Input:} $\theta_0, S, \delta, K, R,\kappa$
\STATE Define $\epsilon = S/n^{1+\delta}$
\STATE $\vartheta_0 \gets \nabla\phi(\theta_0)$
\FOR{$k \in 1,\ldots,K$}
    \STATE Draw $\mathcal{S}_{k-1} \sim \mathcal S$
    \FOR{$i \in \mathcal S_k$}
        \STATE Initialize: $\boldsymbol{\gamma}_{i0}$
        \FOR{$r \in 1,\ldots,R$}
            \STATE Sample $\boldsymbol{\gamma}_{ir}$ from $p(\boldsymbol{\gamma}_{i} | y_i, \theta_k)$
            \STATE Compute $\nabla_{\theta} \log p(y_i,\boldsymbol{\gamma}_{ir} | \boldsymbol{\Omega}_k)$
        \ENDFOR
        \STATE Compute $\widehat{\nabla f_i}^{(R)}(\theta_k) = \frac{1}{R} \sum_{r=1}^R \nabla_{\theta} \log p(y_i,\boldsymbol{\gamma}_{ir} | \theta_k)$
    \ENDFOR
    \STATE Draw $Z_{k-1} \sim N(0, 2\epsilon I_d)$
    \STATE $\vartheta_k \gets \vartheta_{k-1} - \epsilon \left\{\nabla f_0(\theta_{k-1}) + \frac{n}{S}\sum_{i\in\mathcal{S}_{k-1}} f_i(\theta_{k-1})\right\} + \sqrt{\nabla^2\phi(\theta_k)}Z_{k-1}$
    \STATE $\theta_k\gets\nabla\phi^*(\vartheta_k)$
\ENDFOR
\STATE Compute MCMC sample mean $\widehat{\vartheta^*}$ of $(\vartheta_k)_{k=0}^K$ and $\widehat{\theta^*} = \nabla\phi^*(\widehat{\vartheta^*})$
\STATE Compute MCMC sample variance $\widehat{\Sigma}$ of $(\vartheta_k)_{k=0}^K$
\STATE Compute an estimate of matrix $\Psi^{(R)}(\theta^*)$ \COMMENT{Formula is given in \eqref{eq:cov_grad_full_est}}
\STATE Computae an estimate of matrix $\Gamma^{(R)}(\vartheta^*) = \epsilon n^2\Psi^{(R)}(\theta^*) / (2S) + A(\vartheta^*)$
\STATE Solve Lyapunov equation \eqref{eq:lyapunov} to find an estimate of $H^{-1}$
\STATE Obtain corrected $(\vartheta_k^{\text{corr}})_{k=0}^{K}$ \COMMENT{Formula is given in \eqref{eq:rescale}}
\FOR{$k \in 1,\ldots,K$}
    \STATE $\theta_k^{\text{corr}}\gets \nabla\phi^*(\vartheta_k^{\rm corr})$
\ENDFOR
\end{algorithmic}
\end{algorithm}

\section{Proofs}

\subsection{Proof of Theorem 4.4}
\label{sec:proof1}

We assume the readers are familiar with stochastic calculus results as expounded by Chapters 4--5 \cite{le2016brownian}.
A note on notation: let $B_t = (B_t^1,\ldots,B_t^d)^\top$ be a $\R^d$-valued Brownian motion, and $X_t$ adapted to the filtration generated by $B_t$. 
For $A:\R^d\to\R^{d\times d}$ a positive definite matrix-valued continuous map (with a positive square root $\sqrt{A}$), we write
\begin{equation*}
    \int_0^t\sqrt{A(X_s)}\cdot dB_s := 
    \left(
    \sum_{j=1}^{d}\int_0^t\left(\sqrt{A(X_s)}\right)_{1,j}dB_s^1,\;\ldots,\;
    \sum_{j=1}^{d}\int_0^t\left(\sqrt{A(X_s)}\right)_{d,j}dB_s^d
    \right)^\top.
\end{equation*}

By triangle inequality,
\begin{equation}
    \label{eq:triangle1}
    W_2(\rho_k,\nabla\phi_\#\pi) \leq 
    W_2(\rho_k^{\rm MLA}, \nabla\phi_\#\pi) + W_2(\rho_k,\rho_k^{\rm MLA}),
\end{equation}
where $\rho_k^{\rm MLA}$ is the marginal distribution of the $k$-th iterate initialized at the same $\vartheta_0$ and evolving according to the mirror Langevin algorithm \eqref{eq:mla} with no data subsampling. 

\noindent {\bf Sampling error of discretizing MLD.}
We will first derive a sampling error bound assuming no data subsampling, which is quantified by the first term in \eqref{eq:triangle1}.
Set $\rho_0 = \rho_0^{\rm MLA}$ and assume that $k\geq 1$ and that $\rho_{k-1}^{\rm MLA}$ and $\nabla\phi_\#\pi$ are optimally coupled, which is always possible, because the infimum in the definition of a Wasserstein distance is always attained
(Theorem 2.10 \cite{ambrosio2021lectures}).
Denote by $(\vartheta_{k-1}^{\rm MLA},\vartheta_0^{\rm SDE})$ the coupled random variables, and consider the following dynamics, coupled through the same Brownian motion $\{B_t\}_{0\leq t\leq\epsilon}$:
\begin{equation*}
    \begin{cases}
    \vartheta_k^{\rm MLA} = \vartheta_{k-1}^{\rm MLA} - 
    \int_0^\epsilon V(\vartheta_{k-1}^{\rm MLA})~dt + \sqrt{2}\int_0^{\epsilon}\sqrt{A(\vartheta_{k-1}^{\rm MLA})}\cdot dB_t;\\
    \vartheta_\epsilon^{\rm SDE} = \vartheta_{0}^{\rm SDE} - \int_{0}^\epsilon V(\vartheta_t^{\rm SDE})~dt + \sqrt{2}\int_0^\epsilon \sqrt{A(\vartheta_t^{\rm SDE})}\cdot dB_t.
    \end{cases}
\end{equation*}
It is clear that the marginal distributions of $\vartheta_k^{\rm MLA}$ and $\vartheta_\epsilon^{\rm SDE}$ are then $\rho_k^{\rm MLA}$ and $\nabla\phi_\#\pi$, respectively.
Let us consider an auxiliary process, $\{\bar{\vartheta}_t^{\rm SDE}\}_{0\leq t\leq\epsilon}$, 
such that $\bar{\vartheta}_0^{\rm SDE} = \vartheta_{k-1}^{\rm MLA}$ and also $\bar{\vartheta}_\epsilon^{\rm SDE} = \bar{\vartheta}_{0}^{\rm SDE} - \int_{0}^\epsilon V(\bar{\vartheta}_t^{\rm SDE})~dt + \sqrt{2}\int_0^\epsilon \sqrt{A(\bar{\vartheta}_t^{\rm SDE})}\cdot dB_t$,
for the same instance of Brownian motion $B_t$.
Denote by $\bar{\rho}_{k,t}$ its marginal distribution.
The goal is to bound 2-Wasserstein distances between each pair of the three processes and use the triangle inequality.
Under Assumptions \ref{assump:selfconcordance} and \ref{assump:strong_logconcavity}, by Lemma 4 \cite{li2022mirror},
we have
\begin{equation}
    \label{eq:contractive_sde}
    \E[\|\vartheta_{\epsilon}^{\rm SDE} - \bar{\vartheta}_{\epsilon}^{\rm SDE}\|^2|\vartheta_{k-1}^{\rm MLA},\vartheta_0^{\rm SDE}]
    \leq e^{-2(nm-\alpha)\epsilon}\|\vartheta_{k-1}^{\rm MLA} - \vartheta_0^{\rm SDE}\|^2,
\end{equation}
so that by the optimal coupling assumption, 
\begin{equation}
\label{eq:contractive_bound}
W_2^2(\bar{\rho}_{k,\epsilon},\nabla\phi_\#\pi)\leq e^{-2(nm-\alpha)\epsilon}W_2^2(\rho_{k-1}^{\rm MLA},\nabla\phi_\#\pi),
\end{equation}
and the exponent reduces to $-nm\epsilon$.
Next, we compare the distributions of $\vartheta_k^{\rm MLA}$ and $\bar{\vartheta}_\epsilon^{\rm SDE}$. We have
\begin{align*}
    W_2^2(\bar{\rho}_{k,\epsilon},\rho_k^{\rm MLA}) &\leq \E\|\bar{\vartheta}_t^{\rm SDE} - \vartheta_k^{\rm MLA}\|^2\\
    &\leq \E\left[\left\|\int_0^\epsilon \{V(\bar{\vartheta}_t^{\rm SDE}) - V(\bar{\vartheta}_0^{\rm SDE})\}~dt + 
    \sqrt{2}\int_0^\epsilon \left\{\sqrt{A(\bar{\vartheta}_t^{\rm SDE})} - \sqrt{A(\bar{\vartheta}_0^{\rm SDE})}\right\}\cdot dB_t
    \right\|^2\right]\\
    &\leq \epsilon\int_0^\epsilon\E\|V(\bar{\vartheta}_t^{\rm SDE}) - V(\bar{\vartheta}_0^{\rm SDE})\|^2~dt +
    2\int_0^\epsilon\left\|\sqrt{A(\bar{\vartheta}_t^{\rm SDE})} -\sqrt{A(\bar{\vartheta}_0^{\rm SDE})}\right\|^2_{HS}~dt\\
    &\leq \{(n+1)^2L^2\epsilon + 2\alpha\}\int_0^\epsilon\|\bar{\vartheta}_t^{\rm SDE} - \bar{\vartheta}_0^{\rm SDE}\|^2~dt.
\end{align*}
In the third inequality, we first expanded the square and used the moment property of a stochastic integral and the fact that the quadratic covariation (or bracket, as referred to in Chapter 4 \cite{le2016brownian}) between processes $t$ and $B_t$ is zero.
Since it is assumed $\epsilon\gg 1/n^2$, the ``$+\alpha$'' term in the coefficient can be absorbed into $O((n+1)^2\epsilon^2)$. Setting thus $(n+1)^2L^2\epsilon\geq2\alpha$, and using Lemma \ref{lem:growth_bound} below to bound the integral, we have
\begin{equation*}
    \E\|\bar{\vartheta}_t^{\rm SDE} - \bar{\vartheta}_k^{\rm MLA}\|^2 \leq 2(n+1)^2L^2\epsilon\left(
    2\epsilon^3\E\|V(\bar{\vartheta}_0^{\rm SDE})\|^2 + 6\epsilon^2\E\left\|\sqrt{A(\bar{\vartheta}_0^{\rm SDE})}\right\|^2_{HS}\right).
\end{equation*}
Denoting by $\vartheta\sim\nabla\phi_\#\pi$ the optimally coupled random variable with $\bar{\vartheta}_0^{\rm SDE} = \vartheta_{k-1}^{\rm MLA}$, and $\vartheta^*$ the MLE in the interior of $\Theta^*$.
Repeatedly using triangle inequality and Assumptions \ref{assump:selfconcordance}, \ref{assump:strong_logconcavity}:
\begin{align*}
    &\E\|\bar{\vartheta}_t^{\rm SDE} - \bar{\vartheta}_k^{\rm MLA}\|^2 \leq 2(n+1)^2L^2\epsilon\times\\
    &\quad\quad\bigg\{
    4(n+1)^2L^2\epsilon^3\E\|\bar{\vartheta}_0^{\rm SDE} - \vartheta\|^2 + 
    8(n+1)^2L^2\epsilon^3\E_\pi\|\vartheta - \vartheta^*\|^2 +\\
    &\quad\quad12\epsilon^2\alpha
    \E\left\|\bar{\vartheta}_0^{\rm SDE} - \vartheta\right\|^2 + 24\alpha\epsilon^2\E_\pi\|\vartheta-\vartheta^*\|^2 + 24\alpha\epsilon^2\left\|\sqrt{A(\vartheta^*)}\right\|_{HS}^2\bigg\}.
\end{align*}
To conclude, by sub-additivity of square roots:
\begin{equation}
    \label{eq:diffusion_bound_conclusion}
    \begin{gathered}
    W_2(\bar{\rho}_{k,\epsilon},\rho_k^{\rm MLA}) \leq
    \{3(n+1)^2L^2\epsilon^2 + 5(n+1)L\sqrt{\alpha}\epsilon^{3/2}\}
    W_2(\rho_{k-1},\nabla\phi_\#\pi) +
    C_n\epsilon^2 + D_n\epsilon^{3/2},
    \end{gathered}
\end{equation}
with $C_n := 4(n+1)^2L^2\sqrt{\E_\pi\|\vartheta-\vartheta^*\|^2}$ and $D_n := 7\sqrt{\alpha}(n+1)L\left(\sqrt{\E_\pi\|\vartheta-\vartheta^*\|^2} + \sqrt{\|\sqrt{A(\vartheta^*)}\|_{HS}^2}\right)$.

We now put together \eqref{eq:contractive_bound} and \eqref{eq:diffusion_bound_conclusion}.
We set $nm\geq2\alpha$, and since $\epsilon\ll 1/n$, we can also set $n$ sufficiently large such that $3(n+1)^2L^2\epsilon^2 + 5(n+1)L\sqrt{\alpha}\epsilon^{3/2} + \exp(-nm\epsilon/2) \leq \exp(-nm\epsilon/4)$. Simplifying by $n+1\leq 2n$, we conclude, with recursion in $k$:
\begin{equation}
    \label{eq:bound1}
    W_2(\rho_k^{\rm MLA},\nabla\phi_\#\pi) \leq e^{-knm\epsilon/4}W_2(\rho_0,\nabla\phi_\#\pi) + \frac{56\sqrt{\alpha}L}{m}\epsilon^{1/2}\left(\sqrt{\E_\pi\|\vartheta-\vartheta^*\|^2} + \sqrt{\|\sqrt{A(\vartheta^*)}\|_{HS}^2}\right).
\end{equation}
Under the stated assumptions on $\vartheta^*$ and $\pi$, the coefficient of the second summand is $O(1)$.

\noindent {\bf Pure data subsampling error.}
We now control the second term in \eqref{eq:triangle1}. 
At $k\geq 1$, consider two Markov chains corresponding to the MLA and SMLD, respectively, with the same initial laws $\rho_0$ and coupled with the same Gaussian noise $Z_{k-1}$, as follows:
\begin{equation*}
    \begin{cases}
    \vartheta_k^{(1)} = \vartheta_{k-1}^{(1)} - 
    \epsilon V(\vartheta_{k-1}^{(1)}) + \sqrt{2\epsilon A(\vartheta_{k-1}^{(1)})} Z_{k-1};\\
    \vartheta_k^{(2)} = \vartheta_{k-1}^{(2)} - 
    \epsilon\left\{\frac{n}{S}\sum_{i\in\mathcal{S}_{k-1}} V_i(\vartheta_{k-1}^{(2)}) + V_0(\vartheta_{k-1}^{(2)})\right\}+ 
    \sqrt{2\epsilon A(\vartheta_{k-1}^{(2)})} Z_{k-1}.
    \end{cases}
\end{equation*}
Consider a random vector $(\vartheta_{k-1}^{(1)},\vartheta_{k-1}^{(2)},\vartheta^{SMLD})\in\Rd\times\Rd\times\Rd$. 
It is a known result that there exists a probability measure such that its pushforward under the projection onto the 1st and 2nd Euclidean coordinates optimally couples the marginal distributions of $\vartheta_{k-1}^{(1)}$ and $\vartheta_{k-1}^{(2)}$,  
while similarly its pushforward under the projection onto the 2nd and 3rd coordinates optimally couples the marginal distribution of $\vartheta_{k-1}^{(2)}$ and the marginal distribution $\pi^{S}$ of $\vartheta^{SMLD}$ (Lemma  8.4, \cite{ambrosio2021lectures}).
Assume that $(\vartheta_{k-1}^{(1)},\vartheta_{k-1}^{(2)},\vartheta^{SMLD})$ has this distribution.
Denote by $\E_k$ the expectation conditioned on the filtration generated by $(\vartheta_j^{(l)})_{1\leq j\leq k,\; 1\leq l\leq 2}$.
By mutual independence between $\mathcal{S}_{k-1}$, $Z_k$ and $(\vartheta_{k-1}^{(1)},\vartheta_{k-1}^{(2)})$, 
and by the unbiasedness property of the stochastic gradient in SMLD, we have
\begin{align*}
    \E_{k-1}\|\vartheta_k^{(1)} - \vartheta_k^{(2)}\|^2 &= 
    \E_{k-1}\|\vartheta_{k-1}^{(1)} - \vartheta_{k-1}^{(2)}\|^2 -
    2\epsilon\E_{k-1}\langle\vartheta_{k-1}^{(1)}-\vartheta_{k-1}^{(2)}, V(\vartheta_{k-1}^{(1)}) - V(\vartheta_{k-1}^{(2)})\rangle\\
    &+ \epsilon^2\E_{k-1}\|V(\vartheta_{k-1}^{(1)}) - V(\vartheta_{k-1}^{(2)})\|^2 + 
    \epsilon^2\E_{k-1}\left\|V(\vartheta_{k-1}^{(2)}) - \frac{n}{S}\sum_{i\in\mathcal{S}_{k-1}}V_i(\vartheta_{k-1}^{(2)}) - V_0(\vartheta_{k-1}^{(2)})\right\|^2\\
    &+ 2\epsilon\E_{k-1}\left\|\sqrt{A(\vartheta_{k-1}^{(1)})} - \sqrt{A(\vartheta_{k-1}^{(2)})}\right\|_{HS}^2.
\end{align*}
The summand of the last line is bounded above by $2\epsilon\alpha\|\vartheta_{k-1}^{(1)} - \vartheta_{k-1}^{(2)}\|^2$ by Assumption \ref{assump:selfconcordance}.
For the rest of the terms, the bounds follow from the proof of Lemma 8.(ii) \cite{brosse2018promises}: 
namely, from Assumptions \ref{assump:smoothness} and \ref{assump:strong_logconcavity}, one can use the fact that $V$ is $(n+1)L$-co-coercive \cite{zhu1996co}, and thus
\begin{equation}
    \label{eq:cocoercive}
    \epsilon^2\E_{k-1}\|V(\vartheta_{k-1}^{(1)}) - V(\vartheta_{k-1}^{(2)})\|^2 \leq
    (n+1)\epsilon^2L
    \langle\vartheta_{k-1}^{(1)}-\vartheta_{k-1}^{(2)}, V(\vartheta_{k-1}^{(1)}) - V(\vartheta_{k-1}^{(2)})\rangle;
\end{equation}
then, using Assumption \ref{assump:strong_logconcavity} once again and setting $\epsilon \leq \frac{1}{(n+1)L}$, we obtain
\begin{align*}
    \E_{k-1}\|\vartheta_k^{(1)} - \vartheta_k^{(2)}\|^2 
    &\leq \{1 - (nm-2\alpha)\epsilon\}\|\vartheta_{k-1}^{(1)} - \vartheta_{k-1}^{(2)}\|^2 + \epsilon^2\E_{k-1}
    \left\|V(\vartheta_{k-1}^{(2)}) - \frac{n}{S}\sum_{i\in\mathcal{S}_{k-1}}V_i(\vartheta_{k-1}^{(2)}) - V_0(\vartheta_{k-1}^{(2)})\right\|^2\\
    &= \{1 - (nm-2\alpha)\epsilon\}\|\vartheta_{k-1}^{(1)} - \vartheta_{k-1}^{(2)}\|^2+
    \frac{\epsilon^2n}{S}\sum_{i=1}^{n}\left\|V_i(\vartheta_{k-1}^{(2)}) - \frac{1}{n}\sum_{i=1}^{n}V_i(\vartheta_{k-1}^{(2)})\right\|^2.
\end{align*}
By Assumption \ref{assump:smoothness} applied to the function $\vartheta\mapsto V_i(\vartheta) - \frac{1}{n}\sum_{i=1}^{n}V_i(\vartheta)$ for each $i\in[n]$, we obtain:
\begin{align*}
    \E\|\vartheta_k^{(1)}-\vartheta_k^{(2)}\|^2 
    &\leq \{1-(nm-2\alpha)\epsilon\}
    \E\|\vartheta_{k-1}^{(1)} - \vartheta_{k-1}^{(2)}\|^2\\
    &+\frac{\epsilon^2n}{S}\sum_{i=1}^{n}
    \bigg\{
    8L^2\E\|\vartheta_{k-1}^{(2)} - \vartheta^{\rm SMLD}\|^2  + 16L^2\E\|\vartheta^{\rm SMLD} - \vartheta^*\|^2\\
    &\quad\quad +16L^2\left\|V_i(\vartheta^*) - \frac{1}{n}\sum_{i=1}^{n}V_i(\vartheta^*)\right\|^2
    \bigg\};
\end{align*}
so that by the optimal coupling assumption,
\begin{align*}
    W_2^2(\rho_k^{\rm MLA},\rho_k)&\leq \E\|\vartheta_k^{(1)}-\vartheta_k^{(2)}\|^2 \leq \{1-(nm-2\alpha)\epsilon\}W_2^2(\rho_{k-1}^{\rm MLA},\rho_{k-1})+\\
    &\quad\quad\frac{\epsilon^2n}{S}
    \left\{
    8nL^2W_2^2(\rho_{k-1},\pi^{\rm S}) + 16nL^2\E_{\pi^{\rm S}}\|\vartheta^{\rm SMLD} - \vartheta^*\|^2 + 16L^2\sum_{i=1}^{n}
    \left\|V_i(\vartheta^*)\right\|^2
    \right\}.
\end{align*}
Lemmas \ref{lem:ergodic_smld} and \ref{lem:taylor2} (specialized to $\rho_0 = \pi^{\rm S}$ and arbitrary $k$) can be used to control the terms $W_2^2(\rho_{k-1},\pi^{\rm S})$ and $\E_{\pi^{\rm S}}\|\vartheta^{\rm SMLD} - \vartheta^*\|^2$, respectively.
Setting $nm\geq4\alpha$ and recursion in $k$ yields
\begin{equation}
    \label{eq:bound2}
    W_2^2(\rho_k^{\rm MLA},\rho_k)\leq\frac{2\epsilon n}{mS}
    \left\{8L^2W_2^2(\rho_0,\pi^{\rm S}) + \frac{64L^2\epsilon n}{Sm}I_n + 16L^2J_n + \frac{128}{nm}\left\|\sqrt{A(\vartheta^*)}\right\|_{HS}^2\right\},
\end{equation}
where $I_n := \frac{1}{n}\sum_{i=1}^{n}\|V_i(\vartheta^*) - \frac{1}{n}\sum_{j=1}^{n}V_j(\vartheta^*)\|^2$ and $J_n = \frac{1}{n}\sum_{i=1}^{n}\|V_i(\vartheta^*)\|^2$.
Both terms are $O(1)$ under the stated assumption on $\vartheta^*$.

The conclusion of the theorem is now obtained by putting together \eqref{eq:bound1} and \eqref{eq:bound2} and returning to \eqref{eq:triangle1}.

\subsection{Supporting Results for Theorem 4.4}

\begin{lemma}[Growth bound for a mirror diffusion] \label{lem:growth_bound}
    Suppose $\{\vartheta_s\}_{0\leq s\leq t}$ evolves from a fixed $\vartheta_0$ according to the mirror Langevin dynamics \eqref{eq:mld}. Whenever $t\leq \frac{1}{2(n+1)L}$ and 
    $n\geq (4\alpha+1)/2L$, we have
\begin{equation*}
    \E\|\vartheta_0 - \vartheta_t\|^2 \leq 6t^2\E\|V(\vartheta_0)\|^2 + 
    12t\E\left\|\sqrt{A(\vartheta_0)}\right\|_{HS}^2.
\end{equation*}
\end{lemma}
\begin{proof}
We have
\begin{align*}
    \E\|\vartheta_0- \vartheta_t\|^2 &\leq 
    t\int_0^t \E\|V(\vartheta_s)\|^2~ds +
    2\int_0^t\E\left\|A(\vartheta_s)\right\|_{HS}^2~ds\\
    &\leq 2(n+1)^2L^2t\int_0^t\E\|\vartheta_s - \vartheta_0\|^2~ds + 2t^2\E\|V(\vartheta_0)\|^2+4\alpha\int_0^t\E\|\vartheta_s - \vartheta_0\|^2~ds + 4t\E\left\|\sqrt{A(\vartheta_0)}\right\|_{HS}^2;
\end{align*}
by Gr{\"o}nwall's inequality,
\begin{equation*}
    \E\|\vartheta_0 - \vartheta_t\|^2 \leq 
    \left(2t^2\E\|V(\vartheta_0)\|^2 + 4t\E\left\|\sqrt{A(\vartheta_0)}\right\|_{HS}^2\right)\exp\left\{2(n+1)^2L^2t^2 + 4\alpha t\right\},
\end{equation*}
and then the stated conclusion follows by the assumptions on $t$ and $n$.
\end{proof}

\begin{lemma}[Geometric Ergodicity of SMLD]
    \label{lem:ergodic_smld}
    Let $\rho_0\in\mathcal{P}^2(\R^d)$, $\epsilon\leq\frac{1}{(n+1)L}$ and $n\geq 4\alpha/m$. 
    The SMLD algorithm \eqref{eq:smld} has a unique invariant distribution, denoted by $\pi^{\rm S}$, such that $W_2^2(\rho_k,\pi^{\rm S})\leq (1-nm\epsilon/2)^kW_2^2(\rho_0,\pi^{\rm S})$ for $\vartheta_0\sim\rho_0$.
\end{lemma}

\begin{proof}
    Consider two Markov chains $\vartheta_k^{(1)},\vartheta_k^{(2)}$ evolving according to \eqref{eq:smld},
    with $\vartheta_k^{(1)}\sim\rho_k$ and $\vartheta_k^{(2)}\sim\tilde{\rho}_k$ for each $k$.
    We couple with the same Gaussian noise $Z_k$ and same random minibatch of indices $\mathcal{S}_k$ at each iteration:
    \begin{equation*}
    \begin{cases}
    \vartheta_k^{(1)} = \vartheta_{k-1}^{(1)} - 
    \epsilon\left\{\frac{n}{S}\sum_{i\in\mathcal{S}_{k-1}} V_i(\vartheta_{k-1}^{(1)}) + V_0(\vartheta_{k-1}^{(1)})\right\}+ 
    \sqrt{2\epsilon A(\vartheta_{k-1}^{(2)})} Z_{k-1},\\
    \vartheta_k^{(1)} = \vartheta_{k-1}^{(1)} - 
    \epsilon\left\{\frac{n}{S}\sum_{i\in\mathcal{S}_{k-1}} V_i(\vartheta_{k-1}^{(2)}) + V_0(\vartheta_{k-1}^{(2)})\right\}+ 
    \sqrt{2\epsilon A(\vartheta_{k-1}^{(2)})} Z_{k-1}.
    \end{cases}
    \end{equation*}
    We can proceed similarly as in the proof of Lemma 1 \cite{brosse2018promises}. Using the conditional expectation notation $\E_k$ from the proof of Theorem \ref{thm:w2_err} above, we expand the square and use co-coercivity estimate \eqref{eq:cocoercive} and Assumptions \ref{assump:selfconcordance}, \ref{assump:strong_logconcavity} to conclude
    \begin{equation*}
        \E_{k-1}\|\vartheta_k^{(1)} - \vartheta_k^{(2)}\|^2 \leq \{1-(nm-2\alpha)\epsilon\}\|\vartheta_{k-1}^{(1)} - \vartheta_{k-1}^{(2)}\|^2.
    \end{equation*}
    Setting $2\alpha\leq(nm)/2$ and using recursion in $k$, we obtain $W_2^2(\rho_k,\tilde{\rho}_k) \leq (1-nm\epsilon/2)^kW_2^2(\rho_0,\tilde{\rho}_0)$, 
    where $\tilde{\rho}_k$ is the marginal distribution of $\vartheta_k^{(2)}$.
    The conclusion about the unique existence of an invariant measure then follows exactly according to the arguments at the end of Lemma 1 \cite{brosse2018promises} after eqn. (15) therein.
\end{proof}

\begin{lemma}[SMLD second-order moment centered around the MLE]
    \label{lem:taylor2}
    Let $\rho_0\in\mathcal{P}^2(\R^d)$, $\epsilon\leq\frac{1}{2(n+1)L}$ and $n\geq 8\alpha/m$. The SMLD algorithm \eqref{eq:smld} satisfies
    \begin{equation*}
        \E\|\vartheta_k - \vartheta^*\|^2 \leq (1-nm\epsilon/2)^kW_2^2(\rho_0,\pi^{\rm S}) + 
        \frac{4\epsilon}{Sm}\sum_{i=1}^{n}\left\|V_i(\vartheta^*) - \frac{1}{n}\sum_{j=1}^{n}V_j(\vartheta^*)\right\|^2 + \frac{8}{nm}\left\|\sqrt{A(\vartheta^*)}\right\|^2_{HS}.
    \end{equation*}
\end{lemma}

\begin{proof}
    Again using similar arguments as in the preceding lemma and expanding the square, for $\vartheta_k$ initialized at $\vartheta_0\sim\rho_0$ and evolving according to \eqref{eq:smld}, we eventually obtain:
    \begin{align*}
        \E_{k-1}\|\vartheta_k - \vartheta^*\|^2 &\leq \|\vartheta_{k-1} - \vartheta^*\|^2 + 2\epsilon\left\|\sqrt{A(\vartheta_{k-1}})\right\|^2_{HS}\\
        &-2\epsilon\langle\vartheta_{k-1}-\vartheta^*,V(\vartheta_{k-1}) - V(\vartheta^*)\rangle \\
        &+\epsilon^2\E_{k-1}\left[\left\|
        V_0(\vartheta_{k-1}) + \frac{n}{S}\sum_{i\in\mathcal{S}_{k-1}}V_i(\vartheta_{k-1})
        \right\|^2\right].
    \end{align*}
    Isolating the second summand on the right hand side and using triangle inequality and Assumption \ref{assump:selfconcordance} yields
    \begin{equation*}
        2\epsilon\left\|\sqrt{A(\vartheta_{k-1}})\right\|^2_{HS} \leq 4\alpha\epsilon
        \left\|\vartheta_{k-1}-\vartheta^*\right\|^2_{HS} + 4\epsilon\left\|\sqrt{A(\vartheta^*)}\right\|^2_{HS}.
    \end{equation*}
    The rest of the terms can be bounded exactly in the same way as in the proof of Lemma 8.(ii) \cite{brosse2018promises} with an upper bound, when $\epsilon\leq\frac{1}{2(n+1)L}$:
    \begin{equation*}
        (1-nm\epsilon)\|\vartheta_{k-1} - \vartheta^*\|^2 + \frac{2\epsilon^2n}{S}\sum_{i=1}^{n}
        \left\|V_i(\vartheta^*) - \frac{1}{n}\sum_{j=1}^{n}V_j(\vartheta^*)\right\|^2.
    \end{equation*}
    Combining the two displays yields
    \begin{equation*}
        \E_{k-1}\|\vartheta_k - \vartheta^*\|^2 \leq \{1-(nm-4\alpha)\epsilon\}\|\vartheta_{k-1}-\vartheta^*\|^2 + \frac{2\epsilon^2 n}{S}\sum_{i=1}^{n}
        \left\|V_i(\vartheta^*) - \frac{1}{n}\sum_{j=1}^{n}V_j(\vartheta^*)\right\|^2 + 4\epsilon\left\|\sqrt{A(\vartheta^*)}\right\|_{HS}^2.
    \end{equation*}
    Recursion in $k$ yields the stated result.
\end{proof}

\subsection{Proof of Theorem 4.5}
\label{sec:proof2}

Consider $\vartheta_0\sim\pi^{\rm S}$, the invariant distribution of SMLD algorithm \eqref{eq:smld_glmm} (the dependence on a fixed $R$ is implicit).
Taking the equation \eqref{eq:smld_re} and based on a Taylor expansion in the interior of $\Theta^*$, we have:
\begin{equation*}
    \vartheta_1 - \vartheta^* = 
    \vartheta_0 - \vartheta^* - \epsilon HJ^{-1}(\vartheta_0 - \vartheta^*) + \epsilon R_1(\vartheta_0) +\epsilon\xi(\vartheta_0,\mathcal{S}_0) + \sqrt{2\epsilon A(\vartheta^*)}Z_0 + \left\{\sqrt{2\epsilon A(\vartheta_0)} - \sqrt{2\epsilon A(\vartheta^*)}\right\}Z_0,
\end{equation*}
where the Taylor expansion remainder term satisfies, by Assumption \ref{assump:smoothness_str},
\begin{equation}
    \label{eq:remainder1}
    \frac{\|R_1(\vartheta)\|}{\|\vartheta-\vartheta^*\|^2}\leq\frac{(n+1)L}{2}
\end{equation}
for every $\vartheta\in\Theta^*$.
We take the outer product of each $\R^d$-valued random variable on both sides and simplify the terms,
using the fact that $\vartheta_0\sim\pi^{\rm S}$ and the mutual independence between $\vartheta_0,\mathcal{S}_0$ and $Z_0$.
Eventually, we obtain the following equation:
\begin{equation}
    \label{eq:grand_eqn}
    \begin{gathered}
    HJ^{-1}V + VJ^{-1}H = 2\E[A(\vartheta^*)] + 
    \epsilon\E[\xi(\vartheta_0,\mathcal{S}_0)^{\otimes 2}] +
    \epsilon HJ^{-1}VJ^{-1}H +
    \epsilon\E[R_1(\vartheta_0)^{\otimes 2}]
    +2\E\left[\left\{\sqrt{A(\vartheta_0)} - \sqrt{A(\vartheta^*)}\right\}^{\otimes 2}\right] \\
    + 
    2\sqrt{A(\vartheta^*)}\E\left[\left\{\sqrt{A(\vartheta_0)} - \sqrt{A(\vartheta^*)}\right\}^\top\right] + 
    2\E\left[\left\{\sqrt{A(\vartheta_0)} - \sqrt{A(\vartheta^*)}\right\}\right]\sqrt{A(\vartheta^*)}^\top\\
    -\E[\{R_1(\vartheta_0) + \xi(\vartheta_0,\mathcal{S}_0)\}(\vartheta_0 - \vartheta^*)^\top] -
    \E[(\vartheta_0 - \vartheta^*)\{R_1(\vartheta_0) + \xi(\vartheta_0,\mathcal{S}_0)\}^\top]\\
    + \epsilon\E[\{R_1(\vartheta_0) + \xi(\vartheta_0,\mathcal{S}_0)\}
    (\vartheta_0 - \vartheta^*)^\top J^{-1}H] +
    \epsilon\E[HJ^{-1}(\vartheta_0 -\vartheta^*)\{R_1(\vartheta_0) + \xi(\vartheta_0,\mathcal{S}_0)\}(\vartheta_0 - \vartheta^*)^\top].
    \end{gathered}
\end{equation}
For every $\vartheta$, $\xi(\vartheta,\mathcal{S}_0)$ in \eqref{eq:smld_re} is explicitly defined as
$\xi(\vartheta,\mathcal{S}_0) := \sum_{i=1}^{n}V_i(\vartheta) - \frac{n}{S}\sum_{i\in\mathcal{S}_0}V_i(\vartheta)$;
so $\E[\xi(\vartheta,\mathcal{S}_0)|\vartheta] = 0$ and $\E[\xi(\vartheta,\mathcal{S}_0)^{\otimes2}|\vartheta] = \frac{n^2}{S}\Psi^{(R)}(\vartheta)$ (cf. \eqref{eq:cov_grad_full}).
Conditional on $\vartheta_0$, applying another Taylor expansion to $\xi(\vartheta_0,\mathcal{S}_0)$ yields
\begin{equation*}
    \xi(\vartheta_0,\mathcal{S}_0) = \xi(\vartheta^*,\mathcal{S}_0) + \nabla\xi(\vartheta^*,\mathcal{S}_0)(\vartheta_0 - \vartheta^*) + R_2(\vartheta_0),
\end{equation*}
with again $\E[\nabla\xi(\vartheta^*,\mathcal{S}_0)] = 0\in\R^{d\times d}$ and 
\begin{equation}
    \label{eq:remainder2}
    \frac{\|R_2(\vartheta)\|}{\|\vartheta - \vartheta^*\|^2} \leq \frac{(n+1)L}{2}
\end{equation}
for every $\vartheta\in\Theta^*$ by Assumption \ref{assump:smoothness_str}.
Then
\begin{align*}
    \xi(\vartheta_0,\mathcal{S}_0)^{\otimes 2} &= \xi(\vartheta^*,\mathcal{S}_0)^{\otimes 2} + R_3(\vartheta_0),\\
    \R^{d\times d}\ni R_3(\vartheta_0) &= \{\nabla\xi(\vartheta^*,\mathcal{S}_0)(\vartheta_0 - \vartheta^*)\}^{\otimes 2}\\
    &+\xi(\vartheta^*,\mathcal{S}_0)(\vartheta_0-\vartheta^*)^\top\nabla\xi(\vartheta^*,\mathcal{S}_0) + 
    \nabla\xi(\vartheta^*,\mathcal{S}_0)(\vartheta_0-\vartheta^*)\xi(\vartheta^*,\mathcal{S}_0)^\top\\
    &+ \{\xi(\vartheta^*,\mathcal{S}_0) + \nabla\xi(\vartheta^*,\mathcal{S}_0)(\vartheta_0 - \vartheta^*)\}R_2(\vartheta_0)^\top\\
    &+R_2(\vartheta_0)\{\xi(\vartheta^*,\mathcal{S}_0) + \nabla\xi(\vartheta^*,\mathcal{S}_0)(\vartheta_0 - \vartheta^*)\}^\top\\
    &+ R_2(\vartheta_0)^{\otimes 2}.
\end{align*}
Taking expectations on both sides and using mutual independence between $\mathcal{S}_0$ and $\vartheta_0$, therefore,
\begin{align*}
    \E[R_3(\vartheta_0)] &= \frac{n^2}{S}\Psi^{(R)}(\vartheta^*) + \E[\{\nabla\xi(\vartheta^*,\mathcal{S}_0)(\vartheta_0 - \vartheta^*)\}^{\otimes 2}]\\
    &+ \E[\xi(\vartheta^*,\mathcal{S}_0)(\vartheta_0-\vartheta^*)^\top\nabla\xi(\vartheta^*,\mathcal{S}_0)] +
    \E[\nabla\xi(\vartheta^*,\mathcal{S}_0)(\vartheta_0-\vartheta^*)\xi(\vartheta^*,\mathcal{S}_0)^\top]\\
    &+ \E[R_2(\vartheta_0)^{\otimes 2}].
\end{align*}
Now returning to the equality \eqref{eq:grand_eqn}, we can make the following alterations (cf. \eqref{eq:grad_noise_cov}):
\begin{align*}
    \label{eq:eqn_altered}
    \begin{gathered}
    HJ^{-1}V + VJ^{-1}H = 2\Gamma^{(R)}(\vartheta^*)+
    \epsilon HJ^{-1}VJ^{-1}H
    +\epsilon\E[R_1(\vartheta_0)^{\otimes 2}]\\
    \underbrace{+2\E\left[\left\{\sqrt{A(\vartheta_0)} - \sqrt{A(\vartheta^*)}\right\}^{\otimes 2}\right]
    + 
    2\sqrt{A(\vartheta^*)}\E\left[\left\{\sqrt{A(\vartheta_0)} - \sqrt{A(\vartheta^*)}\right\}^\top\right] + 
    2\E\left[\left\{\sqrt{A(\vartheta_0)} - \sqrt{A(\vartheta^*)}\right\}\right]\sqrt{A(\vartheta^*)}^\top}_{(i)} \\
    \underbrace{-\E[R_1(\vartheta_0)(\vartheta_0 - \vartheta^*)^\top] -
    \E[(\vartheta_0 - \vartheta^*)R_1(\vartheta_0)^\top]
    + \epsilon\E[R_1(\vartheta_0)
    (\vartheta_0 - \vartheta^*)^\top J^{-1}H] +
    \epsilon\E[HJ^{-1}(\vartheta_0 -\vartheta^*)R_1(\vartheta_0)(\vartheta_0 - \vartheta^*)^\top]}_{(ii)} \\
    \underbrace{+\epsilon\E[\xi(\vartheta^*,\mathcal{S}_0)(\vartheta_0-\vartheta^*)^\top\nabla\xi(\vartheta^*,\mathcal{S}_0)] +
    \epsilon\E[\nabla\xi(\vartheta^*,\mathcal{S}_0)(\vartheta_0-\vartheta^*)\xi(\vartheta^*,\mathcal{S}_0)^\top]}_{(iii)}\\
    +\epsilon\E[\{\nabla\xi(\vartheta^*,\mathcal{S}_0)(\vartheta_0 - \vartheta^*)\}^{\otimes 2}] + \epsilon\E[R_2(\vartheta_0)^{\otimes 2}].
    \end{gathered}
\end{align*}
Recall the assumption that $\vartheta^*$ remains in the interior of $\Theta^*$ for all $n\geq 1$. In particular, this implies that $2\Gamma^{(R)}(\vartheta^*) = O(\epsilon n^2)$.
We can now estimate all remainder terms in $n$, as follows.
First, by \eqref{eq:remainder1} and \eqref{eq:remainder2},
\begin{equation*}
    \epsilon\E[R_j(\vartheta_0)^{\otimes 2}] = O(\epsilon n^2 \E_{\pi^{\rm S}}\|\vartheta - \vartheta^*\|^4),\; j = 1,2,
\end{equation*}
which under Lemma \ref{lem:taylor4} reduces to $O(\epsilon^3n^4)$. Next,
by Assumption \ref{assump:selfconcordance} and then by Theorem \ref{thm:w2_err},
\begin{equation*}
    (i) = O(\E_{\pi^{\rm S}}\|\vartheta - \vartheta^*\|^2) = O(\epsilon n).
\end{equation*}
By \eqref{eq:remainder1}, Theorem \ref{thm:w2_err}, and Assumption \ref{assump:smoothness_str},
\begin{equation*}
    (ii) = O\left((1+n\epsilon)\epsilon^{3/2}n^{5/2}\right).
\end{equation*}
By Assumption \ref{assump:smoothness_str}, almost surely applicable to the (random) map $\nabla\xi(\cdot,\mathcal{S}_0)$, and Theorem \ref{thm:w2_err},
\begin{equation*}
    (iii) = O(\epsilon^{3/2}n^{5/2}).
\end{equation*}
Finally, again by Assumption \ref{assump:selfconcordance} and Theorem \ref{thm:w2_err},
\begin{equation*}
    \epsilon HJ^{-1}VJ^{-1}H + 
    \epsilon\E[\{\nabla\xi(\vartheta^*,\mathcal{S}_0)(\vartheta_0 - \vartheta^*)\}^{\otimes 2}] = O(\epsilon^2n^3).
\end{equation*}
Under the maintained assumption: $\epsilon = O(n^{-(1+\delta)}),\delta\in(0,1)$, the dominant terms from the above are $(ii)$ and $(iii)$ with an upper bound $O(n^{1-3\delta/2})$ and thus dominated by $2\Gamma^{(R)}(\vartheta^*) = O(n^{1-\delta})$,
which yields our stated result.

\subsection{Supporting Results for Theorem 4.5}

\begin{lemma}[SMLD fourth-order moment centered around the MLE]
    \label{lem:taylor4}
    The invariant distribution of SMLD $\pi^{\rm S}$ satisfies
    \begin{equation*}
        \E_{\pi^{\rm S}}\|\vartheta - \vartheta^*\|^4 \leq C\epsilon^2n^2
    \end{equation*}
    for some $O_n(1)$ constant $C = C(d,\vartheta^*,A,V)$,
    if $n$ is sufficiently large and $\epsilon\leq\frac{1}{Dn}$ for $D = D(L,\alpha)$.
\end{lemma}

\begin{proof}
    For this proof only, we denote by $\zeta_{k-1} := -\xi(\vartheta_{k-1},\mathcal{S}_{k-1})$ for $\xi(\vartheta_{k-1},\mathcal{S}_{k-1})$ from \eqref{eq:smld_re}, 
    and $\zeta^* := -\xi(\vartheta^*,\mathcal{S}_{k-1})$,
    for a more succinct notation.
    Proceeding similarly as in the proof of Lemma \ref{lem:taylor2}, we may expand the quartic power as follows. 
    \begin{align*}
    \|\vartheta_k-\vartheta^*\|^4 &= \bigg\{\|\vartheta_{k-1} - \vartheta^*\|^2 + 2\epsilon\left\|\sqrt{A(\vartheta_{k-1})}Z_{k-1}\right\|^2 + \epsilon^2\|V(\vartheta_{k-1}) + \zeta_{k-1}\|^2\\
    &-2\epsilon\langle V(\vartheta_{k-1}) + \zeta_{k-1}(\vartheta_{k-1}),\vartheta_{k-1}-\vartheta^*\rangle + \\&\sqrt{2\epsilon}\left\langle\vartheta_{k-1}-\vartheta^*,\sqrt{A(\vartheta_{k-1})}Z_{k-1}\right\rangle - (2\epsilon)^{3/2}\left\langle V(\vartheta_{k-1}) + \zeta_{k-1},\sqrt{A(\vartheta_{k-1})}Z_{k-1}\right\rangle\bigg\}^2.
    \end{align*}
    Taking conditional expectations and by Cauchy-Schwarz inequality:
    \begin{align*}
        \E_{k-1}\|\vartheta_k - \vartheta^*\|^4 &= 
        \E_{k-1}
        \bigg[\|\vartheta_{k-1}-\vartheta^*\|^4 + 4\epsilon^2\left\|\sqrt{A(\vartheta_{k-1})}Z_{k-1}\right\|^4 + \epsilon^4\|V(\vartheta_{k-1}) + \zeta_{k-1}\|^4\\
        &+2\epsilon
        \left\langle\vartheta_{k-1}-\vartheta^*,\sqrt{A(\vartheta_{k-1})}Z_{k-1}\right\rangle^2
        + 4\epsilon\left\|\sqrt{A(\vartheta_{k-1})}Z_{k-1}\right\|^2 \\
        &+ 6\epsilon^2\|\vartheta_{k-1}-\vartheta^*\|^2\|V(\vartheta_{k-1}) + \zeta_{k-1}\|^2 - 4\epsilon\|\vartheta_{k-1}-\vartheta^*\|^2\langle V(\vartheta_{k-1}),\vartheta_{k-1}-\vartheta^*\rangle \\
        &+12\epsilon^3\left\|\sqrt{A(\vartheta_{k-1})}Z_{k-1}\right\|^2\|V(\vartheta_{k-1}) + \zeta_{k-1}\|^2
        -8\epsilon^2\left\|\sqrt{A(\vartheta_{k-1})}Z_{k-1}\right\|^2\langle V(\vartheta_{k-1}), \vartheta_{k-1}-\vartheta^*\rangle\\
        &-4\epsilon^3\|V(\vartheta_{k-1}) + \zeta_{k-1}\|^2\langle V(\vartheta_{k-1}) + \zeta_{k-1}, \vartheta_{k-1} - \vartheta^*\rangle\\
        &-8\epsilon^2\left\langle\vartheta_{k-1}-\vartheta^*,
        \sqrt{A(\vartheta_{k-1})}Z_{k-1}\right\rangle\left\langle V(\vartheta_{k-1}) + \zeta_{k-1},
        \sqrt{A(\vartheta_{k-1})}Z_{k-1}\right\rangle\bigg].
    \end{align*}
    We isolate the terms involving the square root of $A(\vartheta_{k-1})$; for the rest of the terms, an upper bound follows exactly by the proof of Lemma 9 \cite{brosse2018promises}.
    In general, if $Z\sim{\rm Normal}_d(0,C)$ for a covariance matrix $C$, then $\E\|Z\|^4 = \sum_{i,j}C_{ii}C_{jj} + 4\sum_{i<j}C_{ij}^2 \leq \left\|\sqrt{C}\right\|_{HS}^4 + 2\|C\|_{HS}^2$.
    The latter term is equal to the $2\|\sqrt{C}\|_4^4$ for $\|\cdot\|_4$ the 4-Schatten norm, so by standard matrix norm inequalities, it is at most $2\left\|\sqrt{C}\right\|_{HS}^4$.
    Based on this fact, using triangle inequalities, we obtain the following bounds under Assumption \ref{assump:selfconcordance}:
    \begin{equation}
            \label{eq:new_bounds_4th}
            \E_{k-1}\left\|\sqrt{A(\vartheta_{k-1})}Z_{k-1}\right\|^4 \leq 24\alpha^2\left\|\vartheta_{k-1} - \vartheta^*\right\|^4 + 24\left\|\sqrt{A(\vartheta^*)}\right\|_{HS}^4.
    \end{equation}
    Also, as shown in the proof of Lemma 9 \cite{brosse2018promises}, 
    under Assumptions \ref{assump:smoothness} and \ref{assump:strong_logconcavity} we have
    \begin{align*}
        \|V(\vartheta_{k-1}) + \zeta_{k-1} - \zeta^*\|^2 &\leq (n+1)L
        \langle\vartheta_{k-1}-\vartheta^*, V(\vartheta_{k-1}) + \zeta_{k-1} - \zeta^*\rangle,\\
        \|V(\vartheta_{k-1}) + \zeta_{k-1} - \zeta^*\|^4 &\leq (n+1)^2L^2\|\vartheta_{k-1}-\vartheta^*\|^2
        \langle\vartheta_{k-1}-\vartheta^*, V(\vartheta_{k-1}) + \zeta_{k-1} - \zeta^*\rangle;
    \end{align*}
    using Assumptions \ref{assump:selfconcordance}--\ref{assump:strong_logconcavity} to simplify, we eventually arrive at
    \begin{align*}
        \E_{k-1}\|\vartheta_{k-1}-\vartheta^*\|^4 
        &\leq
        (1+96\epsilon^2\alpha + 4\epsilon\alpha)\|\vartheta_{k-1}-\vartheta^*\|^4 -
        \E_{k-1}\bigg[\{4\epsilon - 8\epsilon^4(n+1)^2L^2 - 12\epsilon^2(n+1)L - 48\epsilon^3\alpha(n+1)L\}\\
        &\quad\times\|\vartheta_{k-1}-\vartheta^*\|^2
        \langle\vartheta_{k-1}-\vartheta^*, V(\vartheta_{k-1}) + \zeta_{k-1}-\zeta^*\rangle\bigg]\\
        &\quad+ \{8\epsilon^3(n+1)L
        \E_{k-1}[\|\zeta^*\|] + 16\epsilon^2(n+1)L\alpha\}\|\vartheta_{k-1}-\vartheta^*\|^3 \\
        &\quad+\E_{k-1}[\{4\epsilon\|A^{1/2}(\vartheta^*)\|_{HS}^2 + 16\epsilon\alpha + 12\epsilon^2\|\zeta^*\|^2 + 48\epsilon^3\alpha\|\zeta^*\|^2 + 48\epsilon^3\|A^{1/2}(\vartheta^*)\|_{HS}^2(n+1)^2L^2\\
        &\quad+16\epsilon^2(n+1)L\|A^{1/2}(\vartheta^*)\|_{HS}^2\}]\times\|\vartheta_{k-1}-\vartheta^*\|^2 +
        8\epsilon^3\|\zeta^*\|^3\E_{k-1}\|\vartheta_{k-1}-\vartheta^*\|
        \\
        &\quad+96\epsilon^2\|A^{1/2}(\vartheta^*)\|_{HS}^4 +\E_{k-1}[8\epsilon^4\|\zeta^*\|^4 + 16\epsilon\|A^{1/2}(\vartheta^*)\|_{HS}^2 + 48\epsilon^3\|A^{1/2}(\vartheta^*)\|^2\|\zeta^*\|^2].
    \end{align*}
    Now using Assumption \ref{assump:strong_logconcavity} and setting $n\geq (100\alpha)/m,\; \epsilon\leq\frac{1}{2(2L^2+3L+12\alpha)(n+1)}$,
    the first two summands together are at most $(1-nm\epsilon)\|\vartheta_{k-1}-\vartheta^*\|^4$,
    so we can again proceed by recursion in $k$.
    The term involving $\|\vartheta_{k-1}-\vartheta^*\|$ can be bounded using the estimate of $W_2(\rho_{k-1},\vartheta^*)$ from Lemma \ref{lem:taylor2}.
    The term involving $\|\vartheta_{k-1}-\vartheta^*\|^3$ can be bounded above by tediously repeating the arguments just given for a cubic expansion.
    All this combined yields a conclusion for an arbitrary initial distribution $\rho_0$, and then simplifies to $\rho_0 = \pi^{\rm S}$ for arbitrary $k$.
\end{proof}

\subsection{Miscellaneous Proofs}
\label{sec:proof_misc}

Here we prove that the estimator given by \eqref{eq:cov_grad_full_est} is unbiased with respect to the estimand \eqref{eq:cov_grad_full}.

\begin{lemma}
    Using the notation of Section \ref{sec:mirror}, suppose $\gamma_{ir}\stackrel{iid}{\sim} p(\gamma_i|y_i,\theta)$ for $i\in[n]$ and $r\in[R]$.
    Define 
    \begin{equation*}
    \widehat{\Psi}_i^{(R)}(\theta):=\frac{1}{R(R-1)}\sum_{r=1}^R
    \{\nabla\log p(y_i,\gamma_{ir}|\theta) - \widehat{\nabla f}_i^{(R)}(\theta)\}^{\otimes 2}.
    \end{equation*}
    Then the expectation of $\widehat{\Psi}^{(R)}(\theta)$ \eqref{eq:cov_grad_full_est} is equal to $\Psi^{(R)}(\theta)$ \eqref{eq:cov_grad_full} for each $\theta\in\Theta$.
\end{lemma}

\begin{proof}
    From \eqref{eq:cov_grad_full_est} we have:
    \begin{align*}
        \E[\widehat{\Psi}^{(R)}(\theta)] &= \frac{1}{n}\sum_{i=1}^{n}\E[\widehat{\Phi}_i^{(R)}(\theta)] + \frac{\bar{\Psi}_n(\theta)}{n},\\
        \widehat{\Phi}_i^{(R)}(\theta) := 
        &\left\{
        \widehat{f}_i^{(R)}(\theta) - \frac{1}{n}\sum_{j=1}^{n}\widehat{f}_j^{(R)}(\theta)
        \right\}^{\otimes2},\;
        \bar{\Psi}_n(\theta) := \frac{1}{n}\sum_{i=1}^{n}\Psi_i(\theta).
    \end{align*}
    For each $i\in[n]$, we take the expectation of the former stochastic term over a collection of random variables $\{\gamma_{ir}\}_{i\in[n],r\in[R]}$ as follows,
    using the unbiasedness properties of $\widehat{\nabla f}_i^{(R)}$ and $\widehat{\Psi}_i^{(R)}$:
    \begin{align*}
        \frac{1}{n}\sum_{i=1}^{n}\E[\widehat{\Phi}_i^{(R)}(\theta)]
        &= \frac{1}{n}\sum_{i=1}^{n}\E\left[\widehat{\nabla f}_i^{(R)}(\theta)\widehat{\nabla f}_i^{(R)}(\theta)^\top\right] 
        - \E\left[\left\{
        \frac{1}{n}\sum_{i=1}^{n}\widehat{\nabla f}_i^{(R)}(\theta)
        \right\}^{\otimes2}\right]\\
        &= \frac{1}{n}\sum_{i=1}^{n}\nabla f_i(\theta)^{\otimes2} + \bar{\Psi}_n(\theta) - \left\{\frac{1}{n}\nabla f_i(\theta)\right\}^{\otimes 2} + \V\left[\frac{1}{n}\sum_{i=1}^{n}\widehat{\nabla f}_i^{(R)}(\theta)\right]\\
        &= \frac{1}{n}\sum_{i=1}^{n}\left\{\nabla f_i(\theta) - \frac{1}{n}\sum_{i=1}^{n}\nabla f_i(\theta)\right\}^{\otimes 2} + \bar{\Psi}_n(\theta) + \frac{1}{n^2}\sum_{i=1}^{n}\V[\widehat{f}_i^{(R)}(\theta)]\\
        &= \frac{1}{n}\sum_{i=1}^{n}\left\{\nabla f_i(\theta) - \frac{1}{n}\sum_{i=1}^{n}\nabla f_i(\theta)\right\}^{\otimes 2} + \frac{n-1}{n}\bar{\Psi}_n(\theta).
    \end{align*}
    In the third equality, we have used the fact that for every $i\neq j$, $(\gamma_{i1},\ldots,\gamma_{iR})\in\R^{q\times R}$ and $(\gamma_{j1},\ldots,\gamma_{jR})\in\R^{q\times R}$ are mutually independent.
    Combining the fact with the expectation of $\widehat{\Psi}^{(R)}(\theta)$ proves the stated claim.
\end{proof}

%%%%%%%%%%%%%%%%%%%%%%%%%%%%%%%%%%%%%%%%%%%%%%%%%%%%%%%%%%%%%%%%%%%%%%%%%%%%%%%
%%%%%%%%%%%%%%%%%%%%%%%%%%%%%%%%%%%%%%%%%%%%%%%%%%%%%%%%%%%%%%%%%%%%%%%%%%%%%%%

\end{document}